\documentclass[twoside,11pt]{article} %%% Please do not change this class !!!
\usepackage{fancyhdr}    % headings style
\usepackage{graphicx}
\usepackage{rotating}
\usepackage{amsmath,amssymb}

\usepackage{latexsym}

\def\a{\alpha}
\def\b{\beta}
\def\g{\gamma}
 \def\d{\delta}

\def\G{\Gamma}
\def\be{\begin{equation}}
\def\ee{\end{equation}}
\def\vt{\vartheta}

\def\Sigmakin{{}^{\rm k}\Sigma_\alpha}

\def\sqr#1#2{{\vcenter{\hrule height.#2pt\hbox{\vrule width.#2pt
height#1pt \kern#1pt \vrule width.#2pt}\hrule height.#2pt}}}

\newcommand{\dN}{I \! \! H}

\def\Hext{\dN}
\def\Hg{\mathfrak H}
\def\Dg{\mathfrak D}

\begin{document}
 \baselineskip=11pt

 \title{Recent developments in premetric classical
   electrodynamics\hspace{.25mm}\thanks{\,Invited lectures given at
     the 3rd Summer School in Modern Mathematical Physics, 20-31
     August 2004, Zlatibor, Serbia and Montenegro, see \cite{Proc};
     references updated.}\hspace{5pt}$\,^,$\hspace{2pt}\thanks{\,Work
     supported by the Deutsche Forschungsgemeinschaft (DFG), Bonn,
     project HE 528/20-1.}} \author{\bf{Friedrich
     W.~Hehl}\hspace{.25mm}\thanks{\,e-mail address:
     hehl@thp.uni-koeln.de} \\ \normalsize{Inst.\ for Theor.\ Physics,
     University of Cologne, 50923 K\"oln, Germany, and} \\
   \normalsize{Dept.\ of Physics and Astronomy, University of
     Missouri-Columbia,}\\ \normalsize{Columbia, MO 65211, USA}
   \vspace{2mm} \\ \bf{Yakov Itin}\hspace{.25mm}\thanks{\,e-mail
     address: itin@math.huji.ac.il} \\ \normalsize{Institute of
     Mathematics, Hebrew University of Jerusalem and}\\
   \normalsize{Jerusalem College of Engineering, Jerusalem 91904,
     Israel} \vspace{2mm} \\ \bf{Yuri N.\
     Obukhov}\hspace{.25mm}\thanks{\,e-mail address:
     yo@thp.uni-koeln.de} \\ \normalsize{Inst.\ for Theor.\ Physics,
     University of Cologne, 50923 K\"oln, Germany, and} \\
   \normalsize{Dept.\ of Theor.\ Physics, Moscow State University,
     117234 Moscow, Russia} }

\date{}
%\date{24 October 2006, {\it file Zlatibor11b.tex}}

\maketitle

\begin{abstract}
  Classical electrodynamics can be based on the conservation laws of
  electric charge and magnetic flux. Both laws are independent of the
  metric and the linear connection of spacetime. Within the framework
  of such a {\it premetric} electrodynamics --- provided a {\it
    local\/} and {\it linear\/} constitutive law of the vacuum is
  added --- the propagation of electromagnetic waves in the
  geometric-optics limit can be studied. The wave vectors of the wave
  fronts obey a quartic extended Fresnel equation. If one forbids
  birefringence in vacuum, the light cone emerges and Maxwell-Lorentz
  vacuum electrodynamics can be recovered. If minimal coupling of
  electrodynamics to gravity is assumed, then only the gravitational
  potential, i.e., the metric of spacetime, emerges in the
  constitutive law. We discuss recent results within this general
  framework.
\end{abstract}

\medskip


\section{Introduction}

Premetric classical electrodynamics has been consistently formulated
in 2003 in the book of Hehl and Obukhov \cite{Birkbook}. Related
developments can be found in the recent books of Lindell
\cite{Ismobook} and Russer \cite{Russer} and the articles of
Delphenich \cite{Dave1,Dave2}. One of the basic ideas is to formulate
electrodynamics in such a way that the metric of spacetime (that is,
the gravitational potential of Einstein's theory of gravity) doesn't
enter the fundamental laws of electrodynamics. The practicability of
such an approach has been shown in detail in \cite{Birkbook}, see also
the earlier work of Toupin \& Truesdell \cite{T+T,Toupin}, Post
\cite{Post,Postmap}, Kiehn et al.\ \cite{Kiehn}, and Kovetz
\cite{Attay}. In this lecture, we concentrate on the post 2003
development, for earlier results we refer to the extended bibliography
in \cite{Birkbook}.

After the publication of \cite{Birkbook}, premetric electrodynamics
turned out to be a lively and active subject. Mainly the following
aspects were developed (we order them roughly chronologically):
\begin{enumerate}

\item\label{magneticI} Inclusion of {\it magnetic charges\/} in a
  consistent and metric-free way by Kaiser \cite{Kaiser} and by Hehl
  and Obukhov \cite{magnetic}, earlier work had been done by Edelen
  \cite{Edelen}.

\item\label{QHEI} Application of the premetric formalism to the {\it
    Quantum Hall Effect\/} and prediction of the independence of the
  QHE on an external gravitational by Hehl, Obukhov, and Rosenow
  \cite{Rosenow}.

\item\label{nonminimalI} Possible {\it non-minimal coupling\/} of
  gravity to electromagnetism, in particular to the torsion field, by
  Solanki, Preuss, and Haugan \cite{Solanki,Preuss}, see also Hehl
  and Obukhov \cite{gyros}, and, subsequently, by using the formalism
  of \cite{Birkbook}, by Rubilar et al.\ \cite{torsion1} and by Itin
  et al.\ \cite{Itin:2003jp,torsion2}.

\item\label{signatureI} Derivation of the {\it signature of the
    metric\/} of spacetime from the Lenz rule (connected with
  Faraday's induction law) by Itin et al.\ \cite{Annals}.

\item\label{CFJI} The Lorentz-violating {\it CFJ-electrodynamics\/,}
  see Carroll, Field, and Jackiw \cite{CFJ}, was reformulated in the
  premetric formalism by Itin \cite{Yakov1}. It enhances the
  transparency of the CFJ-approach.

\item\label{birefringenceI} New derivation by L\"ammerzahl et al.\
  \cite{lightcone}, see also \cite{forerunner}, of the metric from
  linear premetric electrodynamics: We forbid {\it birefringence\/}
  in vacuum.

\item\label{dimensionI} New insight in the {\it physical dimensions\/}
  of electrodynamics and in the time variability of the vacuum
  impedance and the fine structure constant by Tobar \cite{Tobar} and
  Hehl \& Obukhov \cite{Okun}.

\item\label{skewonI} Effect of the {\it skewon\/} field on light
  propagation in an quite detailed article by Obukhov et al.\
  \cite{skewon}.

\item\label{axionI} Relation between {\it axion\/} electrodynamics,
  see Wilczek \cite{Wilczek87}, and the so-called Post constraint, see
  Hehl and Obukhov \cite{PostCon}; compare the perfect electromagnetic
  conductor (PEMC) of Lindell and Sihvola
  \cite{LindSihv2004a,LindSihv2004b}. Scattering of electromagnetic
  fields at the PEMC.

\item\label{BeigI} Symbol and {\it hyperbolic polynomials\/} of the
  wave equation and its relation to premetric electrodynamics as
  discussed by Beig \cite{Beig}.

\item\label{nonlinearI} {\it Nonlinear\/} electrodynamics by
  Delphenich \cite{DelphenichNonlinear}, see also
  \cite{Yu+GuNonlinear,skewonMex}.

\end{enumerate}

We will now integrate these topics into a short systematic review of
premetric electrodynamics.

\section{Axiom 1: Electric charge conservation}

We base electrodynamics on {\em electric charge conservation}.
The conservation of electric charge was already recognized as
fundamental law during the time of Franklin (around 1750) well before
Coulomb discovered his force law in 1785. Nowadays, at a time, at
which one can catch single electrons and single protons in traps and
can {\em count} them individually, we are more sure than ever that
electric charge conservation is a valid fundamental law of nature.
Therefore matter carries as a {\em primary quality} something called
electric charge which only occurs in positive or negative units of an
elementary charge $e$ (or, in the case of quarks, in $1/3$th of it)
and which can be counted in principle. Thus it is justified to
introduce the physical dimension of charge $q$ as a new and
independent concept.

Mathematically, the first axiom is most conveniently formulated in terms
of exterior calculus with the help of the 3-form $J$ of electrical current.
Its integral over an arbitrary 3-dimensional domain describes the total charge
contained therein. It can be determined by counting the particles carrying
an elementary charge. The latter can be taken as a unit of charge, in the
sense of the theory of dimensions. The conservation of the current reads,
in exterior calculus,
\begin{equation}
dJ = 0.\label{dJ1}
\end{equation}
This law is metric-independent since it is based on a {\it counting\/}
procedure for elementary charges. Given a spacetime foliation, we can
$1+3$ decompose the current $J=-j\wedge d\sigma +\rho$ into the 2-form
of the electric current density $j$ and the 3-form $\rho$ of the
electric charge density.  Then the charge conservation law assumes a
more familiar form of the continuity equation
\begin{equation}
\dot{\rho} + \underline{d}\,j = 0\,, \label{dJ2}
\end{equation}
where the dot denotes the time derivative (Lie derivative with
respect to a suitable vector field) and $\underline{d}$ the
3-dimensional exterior derivative. Both (\ref{dJ1}) and (\ref{dJ2})
can be equivalently formulated in the integral form, see
\cite{Birkbook} for details.

Because of the first axiom and according to a theorem of de Rham, we
can represent the conserved electric current as an exterior derivative
(a generalized ``divergence"), i.e.,
\begin{equation}
J = dH\,.\label{JdH}
\end{equation}
In this way, we naturally introduce the 2-form $H$ of the {\it
  electromagnetic excitation}. Certainly, this quantity is not
uniquely determined from (\ref{JdH}) since the transformation $H
\rightarrow H + d\psi$ evidently preserves the form of (\ref{JdH}).
However, this arbitrariness is eventually fixed in the actual
measurements performed with the help of ideal conductors and
superconductors, which means that the electromagnetic excitation does
have a direct {\it operational\/} significance, see Raith
\cite{Raith}.

The $(1+3)$-decomposition of $H$ is obtained similarly to the decomposition
of the current $J$, thereby introducing the 2-form ${\cal D}$ of the electric
excitation (historical name: ``dielectric displacement") and the 1-form
${\cal H}$ of the magnetic excitation (``magnetic field"):
\begin{equation}
H = -{\cal H}\wedge d\sigma + {\cal D}.\label{HHD}
\end{equation}
Inserting (\ref{HHD}) as well as the $1+3$ decomposition of the
current into (\ref{JdH}), we recover the pair of the
3-dimensional inhomogeneous Maxwell equations
\begin{equation}\label{axiom1}
  dH=J\>\,\,\begin{cases} \hspace{33pt}\underline{d}\,{\cal D}\>=\rho\,,\\
    -\,\dot{\cal D}+\underline{d}\,{\cal H}\>=j\,.
\end{cases}
\end{equation}

Charge conservation is a law that is valid in macro- as well as in
micro-physics. If charge conservation was violated, we could introduce
a 4-form $N$ (one component only!) according to $dJ=N$. L\"ammerzahl,
Macias, and M\"uller \cite{chargenoncons} studied a seemingly more
restricted class of models violating charge conservation.  Such models
can be used as test theories for experiments on the checking of charge
conservation.

\section{Axiom 2: Lorentz force density}

With charge conservation alone, we arrived at the inhomogeneous
Maxwell equations (\ref{JdH}). Now we need some more input for deriving
the homogeneous Maxwell equations. In the purely electric case with a
test charge $q$, we have in terms of components
\begin{equation}\label{Coul}
{F}_a\sim q\,E_a\,,\
\end{equation}
with ${F}$ as force and $E$ as electric field covector.
Generalizing (\ref{Coul}), the simplest relativistic ansatz for
defining the electromagnetic field reads:
\begin{equation}\label{fieldansatz1}
{\rm force\>\> density}\quad\sim\quad {\rm field\>\>strength}
  \times{\rm charge\>\> current\>\> density}\,.
\end{equation}
We know from conventional Lagrangian mechanics that the {\em force}
$\sim\partial L/\partial x^i$ is represented by a {\em covector} with
the dimension of $h/[dx^i]=(h/t,h/\ell)$ (here $h$, $t$, and $\ell$
denote the dimensions of action, time, and length, respectively). @@@
We know from Lagrangian mechanics that the {\em force} $\sim\partial
L/\partial x^i$ is represented by a {\em covector} with the absolute
dimension of action $h$ (here $h$ is {\em not} the Planck constant but
rather only denotes its {\em dimension}).@@@ Accordingly, with the
covectorial force density $f_\a$, the ansatz (\ref{fieldansatz1}) can
be made more precise. The {\em Lorentz force density\/} acting on a
charge density is encoded into Axiom 2:
\begin{equation}\label{Lorentz}
  f_\alpha =(e_\alpha\rfloor F)\wedge J\,.
\end{equation}
Here $e_\alpha$ is an arbitrary, i.e., anholonomic frame or tetrad, a
basis of the tangent space, with $\alpha=0,1,2,3$, and $\rfloor$
denotes the interior product (contraction). The axiom (\ref{Lorentz})
should be read as an operational procedure for defining the
electromagnetic field strength 2-form $F$ in terms of the force
density $f_\alpha$, known {}from mechanics, and the current density
$J$, known {}from charge conservation.  The $1+3$ decomposition of $F$
reads
\begin{equation}\label{Fdecomp}
  F=E\wedge d\sigma+B\,,
\end{equation}
thus defining the electric and the magnetic field strengths $E$ and
$B$ (a differential 1-form and the 2-form, respectively).

\section{Axiom 3: Magnetic flux conservation}

Let us recall that the (four dimensional) electric current 3-form $J$
decomposes as $J=-j\wedge d\sigma +\rho$. Accordingly,
$\int_{C_3}\rho$ represents the electric charge, $\rho$ the electric
charge density, and $j$ the corresponding current density in three
dimensions. Analogously --- apart from the fact that $J$ is a 3-form
and $F$ a 2-form --- $\int_{C_2}B$ represents the magnetic flux and
$B$ the magnetic flux density. Consequently, $E$ has to be interpreted
as the current of the magnetic flux density. We could call $F$ the
(four dimensional) current of magnetic flux.

Why isn't that conventionally presented in textbooks in this way?
There seem to be two reasons: (i) In textbooks the 4-dimensional point
of view is only a supplementary structure which is used eventually in
order to crown the development of electromagnetism, whereas we take,
in the approach presented, the 4-dimensional view on electromagnetism
as fundamental and indispensable right from the beginning. (ii) A
``charge'' density is assumed intuitively to be a 3-form, i.e., one
has to integrate the charge density over a 3-volume in order to find a
net charge. This is a widespread prejudice. However, also a 2-form
like $B$, if integrated over a 2-dimensional surface, can be
considered to be a ``charge'' (here the magnetic flux) and,
accordingly, a 1-form like $E$ as the corresponding ``current
density'' (here the current of the magnetic flux density). Therefore,
Axiom 1 in local form, $dJ=0$, as electric charge conservation, has as
analog --- in local form ---
\begin{equation}\label{axiom2}
  dF=0\>\,\,\begin{cases} \hspace{22pt}\underline{d}\,B\>=0\,,\\
    \dot{B}+\underline{d}\,E\>=0\,,
\end{cases}
\end{equation}
the law of magnetic flux conservation. Since the Faraday induction law
and the sourcelessness of $B$ are consequences of (\ref{axiom2}), this
axiom has a firm experimental underpinning. Note that the induction
law has the form of a continuity equation for the ``charge density''
$B$.

Nevertheless, similar as with the possible violation of electric
charge conservation, there have been numerous attempts to postulate
the existence of magnetic charge according to
\begin{equation}\label{magneticcharge}
  dF=K\>\,\,\begin{cases} \hspace{24pt}\underline{d}\,B\>=\rho_{\rm
      mg}\,,\\ \hspace{2pt}\dot{B}+\underline{d}\,E\>=j_{\rm mg}\,,
\end{cases}
\end{equation}
with the magnetic current density $K=-j_{\rm mg}\wedge
d\sigma+\rho_{\rm mg}$.

It has been shown by Edelen \cite{Edelen}, and more completely by
Kaiser \cite{Kaiser} and two of us \cite{magnetic}, that the law
(\ref{magneticcharge}) can be made to fit into the premetric
approach, provided electric charge conservation is fulfilled.

Clearly, if besides the electric charge $\rho_{\rm el}$, a magnetic
charge $\rho_{\rm mg}$ is present, we have to expect that the
electromagnetic field exerts a force on it. The absolute dimension
$[J]\times[F]$, which enters the Lorentz force density (\ref{Lorentz}),
is {\it electric charge} $\times$ {\it magnetic flux} $=$ {\it
  action}. The same dimension can be found for the expression
$[K]\times[H]$, namely $[K]$ = {\it magnetic flux} and $[H]$ = {\it
  electric charge}. We recognize here a certain reciprocity between
electricity and magnetisms. The revised Axiom 2, taking into account
the linearity of electrodynamics, would then read
\begin{equation}\label{Lorentzmagnetic}
  f_\alpha=\left(e_\alpha\rfloor F\right)\wedge J- (e_\alpha\rfloor
  H)\wedge K\,.
\end{equation}
The necessity of the minus sign will be explained in the next
section. There it will also be shown that the expression for the
energy-momentum current of the electromagnetic field remains the
same, with or without magnetic charge, a quite remarkable fact that
enables us to revise the premetric axiomatics in such a way as to
accommodate the possible existence of magnetic charge.

However, we shouldn't forget that all experiments done so far attest
to the absence of magnetic charge. Hence the experimental situation
leaves no doubt about the absence of magnetic charge in nature. Even
more so, theoretical reasons make such a structure also implausible:
The modified Axiom 2, see (\ref{Lorentzmagnetic}), could no longer
be read as a clear cut operational definition of the electromagnetic
field strength $F$. It would mix with $H$ in a rather confusing way.

In the 19th and the first half of the 20th century, in spite of
Amp\`ere's hypothesis of the electric origin of magnetic effects ---
which today is no longer a hypothesis but rather an experimental fact
--- magnetostatics was built up in analogy to electrostatics.  The
force on the fictitious magnetic charges were then given by $\rho_{\rm
  mg}\wedge{\cal H}$. This coincides with the leading piece of the
last term of the right-hand-side of (\ref{Lorentzmagnetic}). This is
the reason why ${\cal H}$ was called the magnetic field strength in
analogy to the electric field strength $E$, with $\rho_{\rm el}\wedge
E $ as Lorentz force density, even though the magnetic charges were
assumed to be fictitious. We recognize here once more that the
premetric formalism induces transparency into electrodynamics.

\section{Axiom 4: Energy-momentum current}

The field equations of electrodynamics $dH=J$ and $dF=0$ have been set
up. Now we have to turn to the energy-momentum distribution in the
electromagnetic field. Since the Lorentz force density (\ref{Lorentz})
[or (\ref{Lorentzmagnetic})] determines the relation between the
electromagnetic field and {\it mechanics,} this formula must be the
starting point for a discussion of energy and momentum. If we
introduce for the magnetic piece in (\ref{Lorentzmagnetic}) a factor
$\beta$, to be determined later, then
\begin{equation}\label{Lorentzmagnetic'}
  f_\a =\left(e_\alpha\rfloor F\right)\wedge J-\beta\,
  (e_\alpha\rfloor H)\wedge K\,.
\end{equation}
For $K=0$, we recover the case of the absence of magnetic charges. We
assume magnetic charges and substitute the Maxwell equations into
(\ref{Lorentzmagnetic'}):
\begin{equation}\label{inter1}
  f_\alpha = (e_\alpha\rfloor F)\wedge dH - \beta\, (e_\alpha\rfloor
  H)\wedge dF\,.
\end{equation}
Note, however, that this equation is also valid for ordinary
electrodynamics (without magnetic charges) since under those
conditions $dF=0$. In other words, (\ref{inter1}) is valid in both
cases, with or without magnetic charges. The same will be true for the
subsequent formulas.

We integrate partially both terms in (\ref{inter1}), see
\cite{Birkbook}:
\begin{equation}\label{inter2}
  f_\a= d\bigl[\beta\, F\wedge(e_\a\rfloor H)-H\wedge(e_\a\rfloor
  F)\bigr]
  -\beta\,F\wedge d(e_\a\rfloor H)+ H\wedge d(e_\a\rfloor F)\,.
\end{equation}
The expression under the exterior derivative has already the desired
form. We recall the main formula for the Lie derivative of an
arbitrary form $\Phi$, namely $ {\hbox{\it \char'44}}_{e_\alpha}
\Phi=d(e_\alpha\rfloor\Phi) +e_\alpha\rfloor (d\Phi)\,$, see Frankel
\cite{Ted}. This allows us to transform the second part on the
right-hand-side of (\ref{inter2}):
\begin{eqnarray}\label{inter3}
  f_\a= d\bigl[\beta\,F\wedge(e_\a\rfloor H)&-&H\wedge(e_\a\rfloor
  F)\bigr]\nonumber\\ -\beta\,F\wedge ({\hbox{\it \char'44}}_{e_\a}
  H)&+& H\wedge ({\hbox{\it \char'44}}_{e_\a} F)\nonumber\\
  +\beta\,F\wedge e_\a \rfloor (dH)&-& H\wedge e_\a\rfloor(d F)\,.
\end{eqnarray}
The last line can be put into the form
\begin{equation}\label{inter4}
+\beta\,e_\a\rfloor[F\wedge dH]-\beta\,(e_\a\rfloor F)\wedge dH
 -e_\a\rfloor[H\wedge dF]+(e_\a\rfloor H)\wedge dF\,.
\end{equation}
The expressions in the square brackets are 5-forms and vanish. Two
terms are left over, and we find
\begin{eqnarray}\label{inter3a}
  f_\a= d\bigl[\beta\,F\wedge(e_\a\rfloor H)&-&H\wedge(e_\a\rfloor
  F)\bigr]\nonumber\\ -\beta\,F\wedge ({\hbox{\it \char'44}}_{e_\a} H)&+&
H\wedge ({\hbox{\it \char'44}}_{e_\a} F)\nonumber\\
-\beta\,(e_\alpha\rfloor F)\wedge
    dH &+& (e_\alpha\rfloor H)\wedge dF \,.
\end{eqnarray}

With the help of (\ref{inter1}), the last line can be rewritten as
\begin{equation}\label{lastline}
-\b\,f_\a+(1-\b^2)(e_\a\rfloor H)\wedge dF\,.
\end{equation}
Thus, (\ref{inter3a}) reads
\begin{eqnarray}\label{inter3b}
  (1+\b)\,f_\a= d\bigl[\beta\,F\wedge(e_\a\rfloor H)&-&H\wedge(e_\a\rfloor
  F)\bigr]\nonumber\\ -\beta\,F\wedge ({\hbox{\it \char'44}}_{e_\a} H)&+&
H\wedge ({\hbox{\it \char'44}}_{e_\a} F)\nonumber\\
\hspace{30pt}+(1-\b^2)(e_\a\rfloor H)\wedge dF\,.\hspace{-45pt}&&
\end{eqnarray}
As we mentioned above, also this formula is correct with and without
magnetic charges as long as electric charge conservation $dJ=0$ is
taken for granted. {\it Without\/} magnetic charge the last term drops
out because of $dF=0$. Then the choice of $\b=1$ yields the
conventional energy-momentum current of Minkowski. {\it With\/}
magnetic charge the last term disturbs the whole set up. We cannot
find a reasonable energy-momentum current in this case unless we put
$\b=\pm 1$. However, $\b=-1$ would trivialize (\ref{inter3a}) to a
mathematical identity and the Lorentz force density would be lost.
Hence $\b=+1$ is the only reasonable option. In other words, we choose
$\b=1$ in (\ref{Lorentzmagnetic'}) thus arriving at what we displayed
already in (\ref{Lorentzmagnetic}).

Consequently, in both cases, we arrive at the same relation
\begin{equation}\label{fSX}
  f_\alpha = d\,\Sigmakin + X_\alpha\,,
\end{equation}
with the {\em kinematic energy-momentum\/} 3-form of the
electromagnetic field,
\begin{equation}\label{simax}
  ^ {\rm k} \Sigma_\alpha :={\frac 1 2}\left[F\wedge(e_\alpha\rfloor
    H) - H\wedge (e_\alpha\rfloor F)\right]\,\end{equation}
and the force density 4-form
\begin{equation}\label{Xal}
  X_\alpha := -{\frac 1 2}\left(F\wedge {\hbox{\it
        \char'44}}_{e_\alpha}H-H\wedge{\hbox{\it \char'44}}_{e_\alpha}
    F \right)\,.
\end{equation}
For the derivation of the energy-momentum current, we could
alternatively require (\ref{fSX}) right from the beginning. Then the
first line in (\ref{inter3b}) is reasonable because of the exterior
derivative, the second line has to emerge because of the Lie
derivative (in a special frame it is zero). Accordingly, the third
line has to be zero.

The energy-momentum current (\ref{simax}) in premetric exterior
calculus has also been derived by Lindell \& Jancewicz
\cite{IsmoJancewicz}, Segev \cite{Segev}, and Kaiser \cite{Kaiser},
see also the modified derivation by Itin et al.\ \cite{Itin:2003jp,
  Annals}.

The energy-momentum localization, as specified in (\ref{simax}),
represents our Axiom 4. The current $^ {\rm k} \Sigma_\alpha$ is
electric/magnetic reciprocal under the substitutions $H\rightarrow
\!\zeta F\,,\,F\rightarrow -\frac{1}{\zeta}\,H$, with an arbitrary
pseudo-scalar function $\zeta$, as discussed in detail in
\cite{Birkbook}. There it is also shown that $X_\alpha$ vanishes in
specific cases making then $^ {\rm k} \Sigma_\alpha$ to a conserved
quantity.

\section{Premetric electrodynamics and the Maxwell \\equations}

Until now, all our considerations are generally covariant and
metric-free and connection-free. They are valid in flat Minkowskian
and in curved Riemannian spacetime, that is, in special relativity
theory (SR) and in general relativity theory (GR), and even in a
spacetime possibly carrying torsion and/or nonmetricity.  Therefore
Maxwell's equations in the form of (\ref{axiom1}) and (\ref{axiom2})
represent the optimal formulation of the fundamental laws of classical
electrodynamics. Nonminimal couplings, which induce additional
fundamental constants, will be discussed in Sec.\ref{gravity}. In
order to complete the theory, we need a relation between $H$ and $F$,
which we will discuss in the next section.  However, before that we
will have a look at dimensional analysis, today a largely neglected
subject.

In accordance with the general dimensional analysis of Schouten and
Dorgelo, see \cite{SchoutenPhysicists}, and, in particular, of Post
\cite{Post}, the {\em absolute\/} dimension of $J$ is that of a
charge: $[J]=q$ (since the integral of the current over a
3-dimensional spatial domain yields the total charge). Since the
exterior derivative is dimensionless, $[d]=1$, and since the
electromagnetic excitation is given by the inhomogeneous Maxwell
equation (\ref{JdH}), we conclude that the absolute dimension of the
excitation $[H] = q$. Then, by means of (\ref{HHD}), the absolute
dimensions of the electric and magnetic excitations turn out to be
$[{\cal D}]=q$ and $[{\cal H}]=q/t$.  The {\em relative\/} dimensions
are those of their frame components, $[{\cal H}_a]=q/(t\,\ell)$ and
$[{\cal D}_{ab}]=q/\ell^2$, with the spatial indices $a,b =1,2,3$.
These are the ``physical'' dimensions known to physicists and
engineers. Furthermore, denoting the physical dimension of an action
by ${h}$, the Lorentz force equation (\ref{Lorentz}) shows that the
absolute dimension of the electromagnetic field strength 2-form $F$ is
$[F]=h/q=\phi$, that is, action/charge or magnetic flux $\phi$, see
\cite{Birkbook}. Then by (\ref{Fdecomp}), the absolute dimensions of
the electric and magnetic fields $E$ and $B$ are $[E]=\phi/t$ and
$[B]=\phi$, respectively, and the relative dimensions $[E_a]
=\phi/(t\,\ell)$ and $[B_{ab}]=\phi/\ell^2$.

Our analysis does not make use of the metric. Moreover, it is
generally covariant and as such valid in particular in GR and SR.
Furthermore it is valid on a spacetime manifold of arbitrary
dimension. And, on top of that, our considerations do not depend on
any particular choice of the system of physical units. Whatever your
favorite system of units may be, our results will apply to it. In
short: Our dimensional analysis so far is {\it premetric, generally
  covariant, dimensionally independent, and valid for any system of
  units.}

Quite remarkably, in $1 + 2$ spacetime dimensions we can develop the
complete premetric phenomenological theory of the quantum Hall effect
(QHE) that applies to the an electron gas system confined to a
2-dimensional plane \cite{Rosenow}. The crucial observation is that in
$1 + 2$ dimensions, the current is a 2-form, and the natural
constitutive relation, assuming isotropy, is then
\begin{equation}
J = -\,\sigma_{\rm H}\,F.\label{Hall}
\end{equation}
{}From the Maxwell equations (\ref{JdH}) and (\ref{axiom2}) we find
that $\sigma_{\rm H}$ is constant, and the above dimensional analysis
shows that it has the dimension of conductance. The additional
microscopic study \cite{Rosenow} supports these conclusions and
ultimately identifies $1/\sigma_{\rm H}$ with the von Klitzing
constant $R_{\rm K}$.

Since such a phenomenological scheme is premetric, the gravitational
field neither as metric nor as linear connection ever shows up in this
theory. This leads to the prediction that the QHE does not ``feel"
gravity. In particular, this means that the linear Hall resistance
should remain constant in an arbitrary noninertial frame and in any
gravitational field. An experimental verification of this prediction
is highly desirable.

In contrast to the $(1+2)$-dimensional Hall electrodynamics, the
electrodynamical theory in $1 + 3$ spacetime dimensions normally
incorporates the metric of a flat or a curved spacetime via the
constitutive relation $H=H(F)$ between the excitation and the field
strength. In particular, the standard Maxwell-Lorentz electrodynamics
arises when we assume
\begin{equation}
H = \lambda_0{}\,^\star\!F\,.\label{spacetimerel}
\end{equation}
Here the (4-dimensional) Hodge star $\star$ is defined by the
spacetime metric $g_{ij}$ which in Cartesian coordinates reads
$g_{ij}={\rm diag} (c^2, -1, -1, -1)$. The above dimensional analysis
shows that $\Omega_0 := 1/\lambda_0$ has the dimension of resistance.
Usually, it is called the vacuum impedance.

The Maxwell equations (\ref{JdH}) and (\ref{axiom2}), together with
the Maxwell-Lorentz spacetime relation (\ref{spacetimerel}),
constitute the foundations of classical electrodynamics:
\begin{equation}
d\,^\star \!F=\Omega_0\,J\,,\qquad dF=0\,.\label{maxwell1}
\end{equation}
These laws, in the classical domain, are assumed to be of {\it
  universal validity}. Only if vacuum polarization effects of quantum
electrodynamics are taken care of or if hypothetical nonlocal terms
emerge due to huge accelerations, the spacetime relation $H = H(F)$
can pick up corrections yielding a {\it nonlinear\/} law.

We would like to stress that the presence of {\it two\/} fundamental
constants in the Maxwell-Lorentz theory, namely $c$ {\it and}
$\Omega_0$, is, in our opinion, very much underestimated. A simple
application can be given in the context of of the possible variation
of fundamental physical constants.  The possibility of time and space
variations of the fundamental constants is discussed in the literature
both from an experimental and a theoretical point of view, see
\cite{Karshenboim,Honnef,Uzan}, for example.  Of particular interest
are certain indications that the fine structure constant may slowly
change on a cosmological time scale.  Some authors, see Peres
\cite{Peres1,Peres2}, for example, related some experimental evidence
of the variability of the fine structure ``constant'' to the change of
the speed of light $c=c(t,x^a)$.  However, a closer look at the
definition of the fine structure constant
\begin{equation}\label{fine}
\alpha_{\rm f} = {\frac {e^2}{2\varepsilon_0\,c\,{\rm h}}} =
{\frac {e^2}{2\,{\rm h}\,\lambda_0}} = {\frac {\Omega_0}{2R_{\rm K}}}
\end{equation}
shows, see Tobar \cite{Tobar} and \cite{Okun}, that it is explicitly
given in terms of the ratio of two resistances --- vacuum impedance
$\Omega_0$ and von Klitzing constant $R_{\rm K}$ (the quantum Hall
resistance). Note that the speed of light $c$ {\it disappeared
  completely!} Tobar \cite{Tobar} demonstrated explicitly that (27)
doesn't depend on the system of units. This is inbuilt in our
formalism right from its axiomatic beginnings \cite{Birkbook}.

In other words, the formula (\ref{fine}) demonstrates that of the two
fundamental constants of electrodynamics, which appear naturally in
Maxwell-Lorentz electrodynamics (see the previous paragraph), it is
the vacuum imped\-ance that enters the fine structure constant and
{\it not} the speed of light. Accordingly, a variation of the fine
structure constant $\alpha_{\rm f} = \alpha_{\rm f}(t)$ would force us
to conclude that most probably $\lambda_0 = \lambda_0(t)$. A similar
formalism was actually developed by Bekenstein \cite{Bekenstein},
although he inclines to a different physical interpretation of a
variable electron charge $e$. Since a variable $e$ and/or $\rm h$
would yield to charge and/or flux violation, at least one Maxwell
equation had to be given up. Since we consider this undesirable, we
opt for $\lambda_0 = \lambda_0(t)$.

\section{Axiom 5: Local and linear spacetime relation}

A local and linear spacetime relation appears to be a reasonable
physical assumption in the general setting of the axiomatic approach.
Then, the electromagnetic excitation and the field strength are
related by the local and linear constitutive law
\begin{equation}\label{HchiF}
H_{ij}={\frac 1 2}\,\kappa_{ij}{}^{kl}\,F_{kl}\,.
\end{equation}
The constitutive tensor $\kappa$ has 36 independent components.  One
can decompose this object into its irreducible pieces. Obviously, within the
premetric framework, contraction is the only tool for such a decomposition.
Following Post \cite{Postmap}, we can define the contracted tensor of type
$[^1_1]$
\begin{equation}
\kappa_i{}^k := \kappa_{il}{}^{kl}\,,
\end{equation}
with 16 independent components.
The second contraction yields the pseudo-scalar function
\begin{equation}
\kappa := \kappa_k{}^k = \kappa_{kl}{}^{kl}\,.
\end{equation}
The traceless piece
\begin{equation}
\not\!\kappa_i{}^k := \kappa_i{}^k - {\frac 1 4}\,\kappa\,\delta_i^k
\end{equation}
has 15 independent components. These pieces can now be subtracted out
{}from the original constitutive tensor. Then,
\begin{eqnarray}
  \kappa_{ij}{}^{kl} &=& {}^{(1)}\kappa_{ij}{}^{kl} +
  {}^{(2)}\kappa_{ij}{}^{kl} + {}^{(3)}\kappa_{ij}{}^{kl} \\ &=&
  {}^{(1)}\kappa_{ij}{}^{kl} +
  2\!\not\!\kappa_{[i}{}^{[k}\,\delta_{j]}^{l]} + {\frac 1
    6}\,\kappa\,\delta_{[i}^k\delta_{j]}^l.\label{kap-dec}
\end{eqnarray}
By construction, ${}^{(1)}\kappa_{ij}{}^{kl}$ is the totally traceless
part of the constitutive tensor:
\begin{equation}
{}^{(1)}\kappa_{il}{}^{kl} = 0.\label{notrace}
\end{equation}
Thus, we split $\kappa$ according to $36 = 20 + 15 + 1$, and the
$[^2_2]$ tensor ${}^{(1)}\kappa_{ij}{}^{kl}$ is subject to the 16
constraints (\ref{notrace}) and carries $20 = 36 -16$ components.

One may call ${}^{(1)}\kappa_{ij}{}^{kl}$ the principal or the
metric-dilaton part of the constitutive law. Without such a term,
electromagnetic waves are ruled out, see \cite{Birkbook,skewonMex}. We
further identify the two other irreducible parts with a {\it skewon}
and an {\it axion} field, respectively. Conventionally, the skewon and
the axion fields are introduced by
\begin{equation}
\!\not\!S_i{}^j = -\,{\frac 1 2}\!\not\!\kappa_i{}^j,\qquad
\alpha = {\frac 1 {12}}\,\kappa.\label{Salpha}
\end{equation}

The standard Maxwell-Lorentz electrodynamics (\ref{spacetimerel}) arises
when both skewon and axion vanish, whereas
\begin{equation}
{}^{(1)}\kappa_{ij}{}^{kl} = \lambda_0\,\eta_{ij}{}^{kl}.\label{ML}
\end{equation}
Here $\eta_{ijkl} := \sqrt{-g}\,\hat{\epsilon}_{ijkl}$ and $\eta_{ij}{}^{kl}
= \eta_{ijmn}\,g^{mk}g^{nl}$.

Along with the original $\kappa$-tensor, it is convenient to introduce
an alternative representation of the constitutive tensor, see Post
\cite{Post}:
\begin{equation}
\chi^{ijkl} := {\frac 1 2}\,\epsilon^{ijmn}\,\kappa_{mn}{}^{kl}.\label{chikap}
\end{equation}
Substituting (\ref{kap-dec}) into (\ref{chikap}), we find the
decomposition
\begin{equation}
\chi^{ijkl} = {}^{(1)}\chi^{ijkl} + {}^{(2)}\chi^{ijkl}
+ {}^{(3)}\chi^{ijkl}\,,\label{chi-dec}
\end{equation}
again with principal, skewon, and axion pieces. They are determined by
\begin{eqnarray}
{}^{(1)}\chi^{ijkl} &=& {\frac 1 2}\,\epsilon^{ijmn}\,\,{}^{(1)}
\kappa_{mn}{}^{kl},\label{chi1}\\
{}^{(2)}\chi^{ijkl} &=& {\frac 1 2}\,\epsilon^{ijmn}\,\,{}^{(2)}
\kappa_{mn}{}^{kl} = -\,\epsilon^{ijm[k}\!\not\!\kappa_m{}^{l]},\label{chi2}\\
{}^{(3)}\chi^{ijkl} &=& {\frac 1 2}\,\epsilon^{ijmn}\,\,{}^{(3)}
\kappa_{mn}{}^{kl} = {\frac 1 {12}}\,\epsilon^{ijkl}\,\kappa. \label{chi3}
\end{eqnarray}

Using the S-identity and the K-identity derived in \cite{skewonMex}, we
can verify that $^{(2)}\chi$ is {\it skew-symmetric} under the
exchange of the first and the second index pair, whereas $^{(1)}\chi$
is {\it symmetric}:
\begin{equation}
{}^{(2)}\chi^{ijkl} = -\,{}^{(2)}\chi^{klij},\qquad
{}^{(1)}\chi^{ijkl} = {}^{(1)}\chi^{klij}.
\end{equation}

Performing a $(1+3)$-decomposition of covariant electrodynamics
\cite{Birkbook}, we can write $H$ and $F$ as column 6-vectors with the
components built from the magnetic and electric excitation 3-vectors
${\cal H}_a, {\cal D}^a$ and electric and magnetic field strengths
$E_a, B^a$, respectively. Then the linear spacetime relation
(\ref{HchiF}) reads:
\begin{equation}
  \left(\begin{array}{c} {\cal H}_a \\ {\cal D}^a\end{array}\right)
= \left(\begin{array}{cc} {{\cal C}}^{b}{}_a & {{\cal B}}_{ba} \\
{{\cal A}}^{ba}& {{\cal D}}_{b}{}^a \end{array}\right) \left(
\begin{array}{c} -E_b\\  {B}^b\end{array}\right)\,.\label{CR'}
\end{equation}
Here the constitutive tensor is conveniently represented by the
$6\times 6$ matrix
\begin{equation}\label{kappachi}
  \kappa_I{}^K=\left(\begin{array}{cc} {{\cal C}}^{b}{}_a & {{\cal
          B}}_{ba} \\ {{\cal A}}^{ba}& {{\cal D}}_{b}{}^a
    \end{array}\right)\,,\qquad \chi^{IK}= \left( \begin{array}{cc}
      {\cal B}_{ab}& {\cal D}_a{}^b \\ {\cal C}^a{}_b & {\cal A}^{ab}
    \end{array}\right)\,.
\end{equation}
The constitutive $3\times 3$ matrices ${\cal A,B,C,D}$ are constructed
{}from the components of the original constitutive tensor as
\begin{eqnarray}\label{AB-matrix0}
{\cal A}^{ba}&:=& \chi^{0a0b}\,,\qquad
{\cal B}_{ba} := \frac{1}{4}\,\hat\epsilon_{acd}\,
\hat\epsilon_{bef} \,\chi^{cdef}\,,\\
\label{CD-matrix0}
{\cal C}^a{}_b& :=&\frac{1}{2}\,\hat\epsilon_{bcd}\,\chi^{cd0a}\,,\qquad
{\cal D}_a{}^b := \frac{1}{2}\,\hat\epsilon_{acd}
\,\chi^{0bcd}\,.
\end{eqnarray}
If we resolve with respect to $\chi$, we find the inverse formulas
\begin{eqnarray}\label{AB-matrix0'}
\chi^{0a0b} &=& {\cal A}^{ba}\,,\qquad
\chi^{abcd} = \epsilon^{abe}\,\epsilon^{cdf}\,{\cal B}_{fe}\,,\\
\chi^{0abc} &=& \epsilon^{bcd}\,{\cal D}_d^{\ a}\,,\qquad
\chi^{ab0c} = \epsilon^{abd}\,{\cal C}^c_{\  d}\,.\label{CD-matrix0'}
\end{eqnarray}

The contributions of the principal, the skewon, and the axion parts to
the above constitutive 3-matrices can be written explicitly as
\begin{eqnarray}
  {\cal A}^{ab} &=& -\varepsilon^{ab} -
  \epsilon^{abc}\!\not\!S_c{}^0,\label{A}\\ {\cal B}_{ab} &=&\>\;
  \mu_{ab}^{-1} + \hat{\epsilon}_{abc}\!\not\!S_0{}^c,\label{B}\\
  {\cal C}^a{}_b &=&\>\; \gamma^a{}_b\, - (\!\not\!S_b{}^a - \delta_b^a
  \!\not\!S_c{}^c) + \alpha\,\delta_b^a,\label{C}\\ {\cal D}_a{}^b &=&\>\;
  \gamma^b{}_a\, + (\!\not\!S_a{}^b - \delta_a^b \!\not\!S_c{}^c) +
  \alpha\,\delta_a^b. \label{D}
\end{eqnarray}
The set of the symmetric matrices $\varepsilon^{ab}=\varepsilon^{ba}$
and $\mu_{ab}^{-1} = \mu_{ba}^{-1}$ together with the traceless matrix
$\gamma^a{}_b$ (i.e., $\gamma^c{}_c =0$) comprise the principal part
${}^{(1)}\chi^{ijkl}$ of the constitutive tensor.  Usually,
$\varepsilon^{ab}$ is called {\it permittivity\/} tensor and
$\mu^{-1}_{ab}$ {\it reciprocal permeability\/} tensor
(``impermeability'' tensor), since they describe the polarization and
the magnetization of a medium, respectively. The magnetoelectric
cross-term $\gamma^a{}_b$ is related to the Fresnel-Fizeau effects.
The skewon contributions in (\ref{A}) and (\ref{B}) are responsible
for the electric and magnetic Faraday effects, respectively, whereas
skewon terms in (\ref{C}) and (\ref{D}) describe optical activity.
The particular case of the spatially isotropic skewon,
$\!\not\!S_a{}^b= \frac{s}{2}\,\delta_a^b,\; \!\not\!S_0{}^c =0,\;
\!\not\!S_c{}^0=0$, was studied first by Nieves and Pal \cite{NP94}, who
treated $s$ as a third fundamental constant along with the vacuum
impedance $\Omega_0$ and the speed of light $c$.

%%%%%%%%%%%%%%%%%%%%%%%%%%%%%%%%%%%%%%%%%%%%%%%%%%%%%%%%%%%%%%%%%%%
\section{Light propagation in vacuum}
%%%%%%%%%%%%%%%%%%%%%%%%%%%%%%%%%%%%%%%%%%%%%%%%%%%%%%%%%%%%%%%%%%%

Birefringence effects are usually studied in the geometric optics
approximation. It is equivalent to the Hadamard approach, where one
studies the propagation of a discontinuity in the first derivative of
the electromagnetic field.  The basic notions are then the fields of
the {\it wave\/} covector and the {\it ray\/} vector that encode the
information about the propagation of a wave in a spacetime with a
general constitutive relation.

The crucial observation about the surface of discontinuity $S$
(defined locally by a function $\Phi$ such that $\Phi= const$ on $S$)
is that across $S$ the geometric Hadamard conditions are satisfied for
the components of the electromagnetic field and their derivatives:
$[F_{ij}] = 0, \,[\partial_i F_{jk}] = q_i\,f_{jk},\,[H_{ij}] =
0,\,[\partial_i H_{jk}] = q_i\, h_{jk}$. Here $q_i:=\partial_i\Phi$ is
the wave covector. Then using the Maxwell equations (\ref{JdH}) and
(\ref{axiom2}) and the constitutive law (\ref{HchiF}), we find a system
of algebraic equations for the jump functions:
\begin{equation}\label{4Dwave1}
  {\chi}^{\,ijkl}\, q_{j}\,f_{kl}=0 \,,\qquad {\epsilon}^{\,ijkl}\,
  q_{j}\,f_{kl}=0\,.
\end{equation}
Solving the last equation in (\ref{4Dwave1}) by means of
$f_{ij} = q_ia_j - q_ja_i$, which yields the gauge invariance
$a_i\rightarrow a_i+q_i$, we are able to reduce (\ref{4Dwave1})$_1$
eventually to
\begin{equation}\label{finally}
  {\chi}{}^{\,ijkl}\,q_{j}q_ka_l=0\,.
\end{equation}  
This algebraic system, with the mentioned inbuilt gauge invariance,
has a nontrivial solution for $a_i$ only if the determinant of the
matrix on the left hand side vanishes. After removing the gauge
freedom in the determinant \cite{Birkbook}, we arrive at the
4-dimensionally invariant {\it extended Fresnel equation}
\begin{equation} \label{Fresnel}
{\cal G}^{ijkl}(\chi)\,q_i q_j q_k q_l = 0 \,,
\end{equation}
with the fourth rank Tamm-Rubilar (TR) tensor density of weight $+1$
defined by
\begin{equation}\label{G4}
  {\cal G}^{ijkl}(\chi):=\frac{1}{4!}\,\hat{\epsilon}_{mnpq}\,
  \hat{\epsilon}_{rstu}\, {\chi}^{mnr(i}\, {\chi}^{j|ps|k}\,
  {\chi}^{l)qtu }\,.
\end{equation}
It is totally symmetric, ${\cal G}^{ijkl}(\chi)= {\cal G}^{(ijkl)}(\chi)$,
and thus has  35 independent components.

Different irreducible parts of the constitutive tensor (\ref{kap-dec})
contribute differently to the general Fresnel equation. A
straightforward analysis \cite{Birkbook,skewonMex} shows that the axion
piece drops out completely from the TR-tensor, whereas the two
remaining irreducible parts of the constitutive tensor contributes to
(\ref{G4}) as follows:
\begin{equation} \label{propg8}
{\cal G}^{ijkl}(\chi) % = {\cal G}^{ijkl}({}^{(1)}\chi + {}^{(2)}\chi)
= {\cal G}^{ijkl}({}^{(1)}\chi) + {}^{(1)}
\chi^{\,m(i|n|j}\!\not\!S_m^{\ k} \!\not\!S_n^{\ l)}\,.
\end{equation}

Birefringence (or double refraction) is the direct physical
consequence of the Fresnel equation. In this case the quartic wave
covector surface (\ref{Fresnel}) reduces to the pair of second order
light-cones. However, the influence of a skewon field on the Fresnel
wave surface is qualitatively different: The characteristic sign of
the skewon is the emergence of the specific {\it holes in the quartic
  Fresnel surfaces\/} that correspond to the directions in space along
which the wave propagation is damped out completely \cite{skewon}.
This effect is in complete agreement with our earlier conclusion on
the dissipative nature of the skewon field \cite{Birkbook,skewonMex}.

%%%%%%%%%%%%%%%%%%%%%%%%%%%%%%%%%%%%%%%%%%%%%%%%%%%%%%%%%%%%%%%%%%%
\section{No birefringence and the light cone}
%%%%%%%%%%%%%%%%%%%%%%%%%%%%%%%%%%%%%%%%%%%%%%%%%%%%%%%%%%%%%%%%%%%

It has been shown recently \cite{lightcone} that taking the linear
spacetime relation for granted, one ends up at a Riemannian lightcone
provided one forbids birefringence in vacuum, see also \cite{Laem}.
For this the covariant equation (\ref{Fresnel}) is expanded for the
zeroth component $q_0$ of the 4-wave covector $q=(q_0,q_a)$:
 \begin{equation}\label{Fres1}
  { M_0}\,q_0^4+{ M_1}\,q_0^3+{ M_2}
  \,q_0^2+{ M_3}\,q_0+{ M_4}=0\,,
\end{equation}
where the coefficients $M_i$ are homogeneous functions of degree $i$
in the spatial components $q_a$.  Due to Ferrari (1545), the four
solutions of the quartic equation (\ref{Fres1}) can be written as
 \begin{eqnarray}\label{Ferr}
   q_{0(1)}^\uparrow=\hspace{8pt}\sqrt{\a}+\sqrt{\b+\frac{\g}
     {\sqrt\a}}-\d\,,\quad
   &&q_{0(2)}^\uparrow=\hspace{8pt}\sqrt{\a}-\sqrt{\b+\frac{\g}
     {\sqrt\a}}-\d\,,\qquad\\ 
   q_{0(1)}^\downarrow=-\sqrt{\a}+\sqrt{\b-\frac{\g}{\sqrt\a}}-\d\,,\quad
   &&q_{0(2)}^\downarrow=-\sqrt{\a}-\sqrt{\b-\frac{\g}{\sqrt\a}}-\d\,,\qquad
\end{eqnarray}
where $\a,\b,\g,\d$ depend on the coefficients $M_i$ \cite{lightcone}.

Vanishing birefringence means that there is only one future and only
one past directing light cone. There are two possibilities, namely
 \begin{eqnarray}\label{onecone}
   &&q_{0(1)}^\uparrow=q_{0(2)}^\uparrow\,,\qquad q_{0(1)}^\downarrow=
   q_{0(2)}^\downarrow\,,\qquad {\rm i.e.,}\qquad \b=\g=0\,,\\ 
   &&q_{0(1)}^\uparrow=q_{0(1)}^\downarrow\,,\qquad
   q_{0(2)}^\uparrow=q_{0(2)}^\downarrow\,,\qquad {\rm i.e.,}\qquad
   \a=\g=0\,.
\end{eqnarray}
Accordingly, the quartic wave surface in these cases reads 
  \begin{equation}\label{Fres2}
[(q_0-q_0^\uparrow)(q_0-q_0^\downarrow)]^2=0\,,
\end{equation}
 or, explicitly, dropping the square,
   \begin{equation}\label{Fres3}
g^{ij}q_iq_j:=q_0^2+\frac 12\, \frac {M^a}Mq_0q_a+
 \frac 18\, \Big(4\frac{M^{ab}}M-\frac{M^aM^b}{M^2}\Big)q_aq_b=0\,,
\end{equation}
where $M^a, M^{ab}$ are constructed in terms of the components of
$\chi^{ijkl}$, see \cite{Birkbook}.

Since this relation describes an unique light cone, it should be
understood, up to a scalar factor, as an operational definition of a
Riemannian metric.  From the condition of the existence of a unique
solution (or from hyperbolicity, see Beig \cite{Beig}),
eq.(\ref{Fres3}) has to possess two real solutions for any given
$q_a$. As a consequence, the signature of the metric is Lorentzian.
Accordingly, the signature of the metric is a consequence of the
existence of a unique real solution of the Maxwell equations in a
future causal cone for arbitrary sources with compact support.

An alternative treatment of the correlation between the linear
constitutive relations (\ref{HchiF}) and the signature is given by two
of us in \cite{Annals}.  In order to provide a physical interpretation
of the 4-dimensional quantities, we construct their
$(1+3)$-decompositions with a number of free sign factors. For the
current, we take
\begin{equation}\label{Jdec}
  J=i_{\tt T}\,j\wedge d\sigma+i_{\tt S}\,\rho\,
 \end{equation}
 and for the electromagnetic field
 \begin{equation}\label{HFdec}
 H=h_{\tt T}\,{\mathcal H}\wedge d\sigma+h_{\tt S}\,{\mathcal D}\,,
 \qquad F=f_{\tt T}\,E\wedge d\sigma+f_{\tt S}\,B\,.
 \end{equation}
We introduced here the {\bf{T}}ime and {\bf S}pace factors $i_{\tt
  T},i_{\tt S},h_{\tt T},h_{\tt S}f_{\tt T},,f_{\tt S} =\pm 1$.  
% with values {}from the set $\{+1, -1\}$.
For all possible signatures of the 4-dimensional metric, we obtain the
expressions for the electric and magnetic energy densities that
correspond to (\ref{simax}). The metric of a Lorentzian type turns out
to be related to a positive electromagnetic energy density.  This
result does not depend on the values of the sign factors.

We analyzed the $(1+3)$-decompositions of the field equations, and we
derived which sign factors are conventional and which do depend on the
signature.  We find that the electric charge has two possible signs
for all signatures.  For all signatures, we determine the features of
the interactions between charges and between currents.  Only for a
metric with a Lorentzian signature we have ordinary Maxwell-Lorentz
electrodynamics.  In particular, it yields attraction between opposite
charges and repulsion between charges of the same sign: this is {\it
  Dufay's law}.  Also the magnetic force has a correct sign. In
particular, it is responsible for the pulling of a ferromagnetic core
into a solenoid independently of the direction of the current, in
accordance with {\it Lenz's rule.} That is, we show in the metric-free
approach that positive electromagnetic energy density together with
the correct signs in Dufay's and Lenz's rules correspond to a
Lorentzian signature of the metric.
 
%%%%%%%%%%%%%%%%%%%%%%%%%%%%%%%%%%%%%%%%%%%%%%%%%%%%%%%%%%%%%%%%%%%
\section{Axion electrodynamics and the CFJ model}
%%%%%%%%%%%%%%%%%%%%%%%%%%%%%%%%%%%%%%%%%%%%%%%%%%%%%%%%%%%%%%%%%%%

When the skewon is trivial, $\!\not\!S_i{}^j = 0$, the structure of
the electromagnetic theory simplifies greatly. If, furthermore, the
principal part has the form (\ref{ML}), we end up with the linear
constitutive law $H=\lambda_0\,^\star F+\alpha F$. This framework is
called { axion\/} (Maxwell-Lorentz) { electrodynamics,} see Ni
\cite{Ni73,Ni77,Ni84} and Wilczek \cite{Wilczek87}, e.g.:
\begin{equation}\label{MaxAx}
\lambda_0\,d\,^\star F+(d\alpha)\wedge F=J\,,\qquad dF=0\,.
\end{equation} 
It is as if the current $J$ picked up an additional piece depending on
the gradient of the axion field. In tensor calculus, we have for the
inhomogeneous Maxwell equation
$\lambda_0\,\partial_j(\sqrt{-g}\,F^{ij})+\epsilon^{ijkl}
(\partial_j\alpha) F_{kl}=\check{J}^i$, with
$\check{J}^i=\epsilon^{ijkl}J_{jkl}/6$.

As a degenerate special case, we can also consider the {pure
  (``stand-alone'') axion\/} field with $\,^{(1)}\kappa_{ij}{}^{kl} =
\,^{(2)}\kappa_{ij}{}^{kl}=0$. Then,
\begin{equation}\label{standalone1}
  H=\alpha\,F\qquad{\rm or}\qquad\begin{cases} {\mathcal
    H}=-\alpha\,E\,,\\ {\mathcal D}=\hspace{8pt}\alpha\,B\,,\end{cases}
\end{equation}
and the Maxwell equations read
\begin{equation}\label{standalone2}
  (d\alpha)\wedge F=J\qquad{\rm and}\qquad dF=0\,.
\end{equation}
This is a special case of axion electrodynamics, namely (\ref{MaxAx})
with $\lambda_0=0$. Historically, the first person to discuss (and to 
reject) a constant pure axion field was Schr\"odinger \cite{Schroedinger}, 
p.25, penultimate paragraph, and, as a non-constant field, Dicke 
\cite{Dicke64}. The framework (\ref{standalone1}),(\ref{standalone2}) 
in fact corresponds to Tellegen's {\it gyrator\/} 
\cite{Tellegen1948,Tellegen1956/7} and to Lindell \& Sihvola's {\it perfect 
electromagnetic conductor\/} (PEMC) \cite{LindSihv2004a,LindSihv2004b}.

A further specialization of the axion electrodynamics is possible when
the covector $\nu := d\alpha=\nu_i\,dx^i$ has constant components.
In the cosmological context, this yields the Lorentz-violating 
{\it CFJ-electrodynamics\/,} see Carroll, Field, and Jackiw \cite{CFJ}. 

Even though in the extended Fresnel equation (\ref{Fresnel}) the
birefringence effect is independent of the axion field, such a
behavior was discovered in \cite{CFJ}.  In order to resolve this
problem, we consider the standard wave ansatz
 \begin{equation}\label{lp1}
F=f\,e^{i\varphi}\,, 
\end{equation}
where $i$ is the imaginary unit and $\varphi=\varphi(x^k)$, while $f$
is a constant 2-form.  We denote the wave covector as
$q=d\varphi=q_i\,dx^i$.  For the ansatz (\ref{lp1}), the
components of the excitation 2-form for the CFJ case become
\begin{equation}\label{lp2}
\check{H}^{kl}=\chi^{klmn}\,f_{mn}\,e^{i\varphi}\,, 
\end{equation}
with $\check{H}^{kl}:=\epsilon^{klmn}H_{mn}/2$. In contrast to the
Hadamard method, the amplitude of $\check{H}^{kl}$ is not a constant,
even if $f_{mn}$, the amplitude of $F_{mn}$, is a constant.
Substituting (\ref{lp1}) and (\ref{lp2}) into the field equations and
putting the current $J$ to zero, we obtain a system of 8 linear
equations
\begin{eqnarray}\label{lp3}
  &&\epsilon^{ijkl}q_jf_{kl}=0\,,\qquad
  \Big(\chi^{ijkl}q_j-i(\partial_j\chi^{ijkl})\Big)f_{kl}=0
\end{eqnarray}
for 6 independent variables $f_{kl}$.  For the special case appearing
in the CFJ-model, the constitutive tensor involves a principal part
${}^{(1)}\chi^{ijkl}$ like in conventional vacuum electrodynamics and
the variable axion part
${}^{(3)}\chi^{ijkl}=\alpha(x^m)\epsilon^{ijkl}$.  Following the
procedure given in \cite{Birkbook}, we rewrite this equation in the
covariant form
\begin{equation}\label{lp13}
{\cal G}^{ijkl}(\chi)\,q_i q_j q_k
q_l - \chi^{ijkl}(\partial_i\alpha)(\partial_l\alpha)\,q_j q_k = 0\,.
\end{equation}
Substituting here the CFJ constitutive tensor, we obtain
($\nu_i=\partial_i\alpha$)
 \begin{equation}\label{lp14}
   (q_iq^i)^2-(\nu_iq^i)^2+(\nu_i\nu^i)(q_jq^j)=0\,.
\end{equation}
Finally, if we choose $\alpha=\mu\tau$, with $\mu$ as a constant and
$\tau$ as proper time, we find $\nu_i=(\mu,0,0,0)$ and, together with
$q_i=(\omega,{\bf q})$,
  \begin{equation}\label{lp15}
    (\omega^2-{\bf q}^2)^2-\mu^2{\bf q}^2=0\,,
  \end{equation}
which coincides with the CFJ dispersion law \cite{CFJ}. 

In fact we have here two different types of the birefringence effects:
(i) The premetric birefringence is generated by the algebraic
structure of the constitutive tensor. (ii) The CFJ birefringence is
generated by derivatives of the constitutive tensor.

\section{Axiom 6: Splitting of the electric current}

In order to discuss the electrodynamics of continuous media, we need
some further input. The crucial point is as follows: The total current
density is the sum of the two contributions originating ``from the inside"
of the medium (which is interpreted as a bound or material charge) and
``from the outside" (which is free or external charge):
\begin{equation}
  J = J^{\rm mat} + J^{\rm ext}.\label{total}
\end{equation}
Here, the bound electric current inside matter is denoted by
{\em mat} and the external current by {\em ext}. The same notational
scheme is also applied to the excitation $H$, so we have
$H^{\rm mat}$ and $H^{\rm ext}$.

Bound charges and bound currents are inherent characteristics of matter
determined by the medium itself. They only emerge {\it inside} the
medium.  In contrast, free charges and free currents in general appear
outside and inside matter. They can be prepared for a specific purpose
by a suitable experimental arrangement (a beam of charged particles, say,
and scatter them at the medium), or we could study the reaction of a medium
in response to a prescribed configuration of charges and currents,
$J^{\rm ext}$.

Furthermore, we assume that the charge bound by matter fulfills the usual
charge conservation law separately:
\begin{equation}
d\,J^{{\rm mat}} = 0.\label{Axiom6}
\end{equation}
We call (\ref{total}) together with (\ref{Axiom6}) {\it Axiom 6}.  It
specifies the properties of the classical material medium. In view of
the relation $dJ=0$, resulting from the first axiom, the assumption
(\ref{Axiom6}) means that there is no physical exchange (or
conversion) between the bound and free charges. Although the sixth
axiom certainly does not exhaust all possible types of material media,
it is valid for a sufficiently wide class of media.

Analogously to the Maxwell equation (\ref{JdH}), which was derived
{}from the conservation law (\ref{dJ1}), we introduce by means of
(\ref{Axiom6}) the excitation $H^{{\rm mat}}$ as a ``potential'' for
the bound current:
\begin{equation}
J^{{\rm mat}}=d\,H^{{\rm mat}}\,.\label{curexactM}
\end{equation}
The $(1+3)$-decomposition, following the pattern of (\ref{HHD}), yields
\begin{equation}\label{decomexiM}
  H^{{\rm mat}}=  -{\cal H}^{{\rm mat}}\wedge d\sigma + {\cal D}^{{\rm mat}}.
\end{equation}
The conventional names for these newly introduced excitations are
{\em polarization} 2-form $P$ and {\em magnetization} 1-form $M$,
i.e.,
\begin{equation}
  {\cal D}^{{\rm mat}}\equiv -\,P\,,\qquad {\cal H}^{{\rm mat}}\equiv
  M\,.\label{PM}
\end{equation}
The minus sign is chosen in accordance with the usual behavior of
paramagnetic matter. Then, by using this in (\ref{Axiom6}), we find,
in analogy to the inhomogeneous Maxwell equations (\ref{axiom1}),
\begin{equation}
  -\,\underline{d}\,P = \rho^{\rm mat}\, ,\qquad \underline{d}\,M +
  \dot{P} = j^{\rm mat}.\label{dP}
\end{equation}
The identifications (\ref{PM}) are only true up to an exact form.
However, the uniqueness is guaranteed if we require ${\cal D}^{\rm mat}=0$
for $E=0$ and ${\cal H}^{\rm mat}=0$ for $B=0$, see \cite{Birkbook}.

In order to finalize the scheme, we define the {\it external excitation}
\begin{equation}
\Hext := H - H^{\rm mat}\quad\begin{cases} \hspace{3pt}
\Dg := {\cal D} - {\cal D}^{\rm mat} = {\cal D} + P \\
\hspace{3pt}\Hg := {\cal H} - {\cal H}^{\rm mat} = {\cal H} - M
\end{cases}\label{DHe}
\end{equation}
The external excitation $\Hext = (\Hg, \Dg)$ can be understood as an
auxiliary quantity. When we differentiate (\ref{DHe}) and eliminate $dH$
and $dH^{\rm mat}$ by (\ref{JdH}) and (\ref{curexactM}), respectively,
we find, making use of (\ref{total}), the {\em inhomogeneous} Maxwell
equation for matter:
\begin{equation}\label{MaxMat}
d\Hext = J^{\rm ext} \quad\begin{cases} \hspace{24pt}
\underline{d}\,\Dg = \rho^{\rm ext}\,,\\
\hspace{3pt}\underline{d}\,\Hg - \dot{\Dg} = j^{\rm ext}.\end{cases}
\end{equation}
In Maxwell-Lorentz electrodynamics, we obtain from (\ref{DHe}) and the
universal spacetime relation (\ref{spacetimerel}) the expressions 
\begin{eqnarray}
  \Dg &=&\;\,\varepsilon_0\;^{\underline{*}} E\, +\,
  P(E,B)\,,\\ \Hg &=&\hspace{2pt} {\frac 1{\mu_0}}\;^{\underline{*}}B -
  M(B,E)\,.
\end{eqnarray}

The polarization $P(E,B)$ is a functional of the electromagnetic field
strengths $E$ and $B$. In general, it can depend also on the
temperature $T$ and possibly on other thermodynamic variables
specifying the material continuum under consideration; similar remarks
apply to the magnetization $M(B,E)$. The system (\ref{MaxMat}) looks
similar to the Maxwell equations (\ref{axiom1}). However, the
equations in (\ref{MaxMat}) refer only to the external fields and
sources.  We stress that the {\em homogeneous\/} Maxwell equation {\it
  remains valid\/} in its original form.\bigskip

%--------------------------------------------------
\section{Coupling of electrodynamics to gravity}\label{gravity}
%--------------------------------------------------

The coupling between the electromagnetic and the gravitational field
is an age-old problem. It is already related to the first observable
prediction of GR about the bending of light rays of stars in the
gravitational field of the Sun. The electromagnetic and gravitational
effects are of rather different orders of magnitude. However, the
increasing precision of modern experimental techniques gives rise to
the hope that the appropriate form of the coupling can soon be
determined.

In particular, we have in this context two independent but closely
related problems:
\begin{itemize}
\item[(i)] How does the gravitational field of a massive source depend
  on its electric charge?
\item[(ii)] How does the electromagnetic field of a charged massive
  source change when the gravity is ``switched on''?
\end{itemize}
In most cases, the coupling between two fields can be represented by a
specific term in the total action. This term has to respect the {\em
  diffeomorphism\/} invariance related to gravity as well as the {\em
  gauge\/} invariance of electrodynamics. Moreover, it is reasonable
to require the main facts of both theories (the conservation of
energy-momentum, of electric charge, and of magnetic flux) to be
preserved in the modified model.  Although, coordinate and gauge
invariance strongly restrict the variety of admissible coupling terms,
we have still an infinity set of possibilities. Indeed, we can always
construct a polynomial of a chosen coupling term.  This new term can
also serve as an admissible additional piece of the action. Hence we
will restrict ourselves to the admissible {\em parity conserving\/}
coupling terms of the lowest order, see also \cite{gyros}.

Our main result will be that {\em all\/} such lowest order
modifications of the standard Einstein-Maxwell system are completely
embedded in the axiomatic approach to electrodynamics.

%--------------------
\subsection{Coupling of electrodynamics to Einsteinian gravity}
%--------------------
In the framework of GR, the coupling between the electromagnetic
field and gravity is  managed by the electromagnetic  action
itself
\begin{equation}\label{EM-act}
  S(g,F)=-\frac{\lambda_0}{2}\int\, ^\star\!  F\wedge
  F=-\frac{\lambda_0}{4}\int F_{ij}F^{ij}\sqrt{-g}\,d^4x\,.
\end{equation}
Here $g=g_{ij}\,dx^i\otimes dx^j=o_{\a\b}\,\vt^\a\otimes \vt^\b$ is
the metric tensor, whereas the 2-form $F=F_{ij}dx^i\wedge
dx^j/2=F_{\a\b}\vt^\a \wedge \vt^\b/2$ is the electromagnetic field
strength. The typical form of the coupling Lagrangian (\ref{EM-act})
is $L(g,F)\sim (g^{2}\cdot F^2)$.  Here and later, we use the notation
$(\,\_\,\cdot\,\_\,)$ for a summation that is evaluated by contracting
the indices.

If one adds to (\ref{EM-act}) the actions of the gravitational and the
matter fields, then variation with respect to the metric yields the
Einstein field equation (without cosmological constant)\footnote{In
  exterior calculus, we vary with respect to the orthonormal coframe
  $\vt^\a$ and find $\frac 12\,\eta_{\a\b\g}\wedge R^{\b\g}=\frac{8\pi
    G}{c^3}({}^{\tt (em)}\Sigma_\a+{}^{\tt (mat)}\Sigma_\a)$. Here
  $\eta_{\a\b\g}=\,^\star(\vt_\a\wedge\vt_\b\wedge\vt_\g)$, $R^{\a\b}$
  is the curvature 2-form of the Riemannian spacetime, and $G$
  Newton's gravitational constant. An energy-momentum 3-form
  translates into an energy-momentum tensor according to
  $\Sigma_\a=T_\a{}^\b\eta_\b$, with $\eta_\b=\,^\star\vt_\b$.}
\begin{equation}\label{EM-eq1}
  \hbox{Ric}_{ij}-\frac 12 \, Rg_{ij}=\frac{8\pi G}{c^3}( {}^{\tt
    (em)}T_{ij}+{}^{\tt (mat)}T_{ij})\,,
\end{equation}
with the Ricci tensor $ \hbox{Ric}_{ij}:=R_{kij}{}^k$ and the
curvature scalar $R:=g^{ij}\hbox{Ric}_{ij}$. The electromagnetic and
the material energy-momentum tensors act as sources of the
gravitational field.

The Reissner-Nordstr\"om solution of the Einstein-Maxwell equations
\begin{equation}\label{RN-sol}
  ds^2=\left(1-\frac {2m}r+\frac {q^2}{r^2}\right)dt^2- \left(1-\frac
    {2m}r+\frac {q^2}{r^2}\right)^{-1}dr^2-r^2d\Omega^2\,,
\end{equation}
with $m=GM/c^2$ and $q^2=GQ^2/(4\pi\varepsilon_0c^4)$, describes a
mass $M$ with an electric charge $Q$. The electromagnetic field
strength $F$ is involved in (\ref{EM-act}) only algebraically, that
is, without derivatives. Hence the inhomogeneous Maxwell equation can
only be derived by a variation taken with respect to the potential
$A$.

Let us turn then to a discussion of the potential. In the conventional
textbook approach, see Landau and Lifshitz \cite{LL} and Misner,
Thorne, and Wheeler \cite{MTW}, the {\sl components} $A_i$ of the
electromagnetic potential 1-form $A=A_i\,dx^i$ are taken as field
variables. The recipe for the ``minimal'' transition from SR to GR is
then the ``comma goes to semicolon rule'' \cite{MTW}; the comma ``,''
denotes the partial, the semicolon ``;'' the covariant derivative. For
the {\sl components} $F_{ij}$ of the field strength $F$ we find then
\begin{equation}\label{El-Pot}
  F_{ij}=A_{j;i}-A_{i;j}=A_{j,i}-A_{i,j}\,,
\end{equation}
since the Levi-Civita connection is symmetric. The relation between
the components $F_{ij}$ and $A_i$ is then eventually recognized as
independent of the metric structure. Needless to say that such a
procedure is highly coordinate dependent. 

In contrast, in our axiomatic approach the potential 1-form $A$, as
(coordinate independent) {\sl geometrical object}, is the field
variable of our choice. From magnetic flux conservation we find
directly
\begin{equation}\label{El-Pot'}
 F=dA\,,
\end{equation}
without messing around with covariant derivatives. In this
coordinate-independent way we recognize right away that there is no
chance for the metric to intervene in $F=dA$. Clearly, a coordinate
{\em in\/}dependent procedure is to be preferred against a coordinate
dependent one.

The conventional story then continues as follows: We ``semicolonize''
the Maxwell equations in order to make them fit for survival in the
curved pseudo-Riemannian spacetime of GR,
\begin{equation}\label{Max-eq}
\begin{cases} \hspace{3pt}
F_{ij;k}+F_{jk;i}+F_{ki;j}=0\,,\\
\hspace{3pt} F^{ij}{}_{;j}=\Omega_0\,J^i\,.
 \end{cases}
\end{equation}
In the first equation the metric tensor drops out, again because of
the symmetry of the Levi-Civita connection:
\begin{equation}\label{Max-eq'}
F_{ij,k}+F_{jk,i}+F_{ki,j}=0\,.
\end{equation}
It is not sensitive to the gravity field of the source.  The second
equation can be rewritten as
\begin{equation}\label{Max-eq3}
  (\sqrt{-g}g^{i[k}g^{l]j}F_{kl})_{,j}=\Omega_0\,\sqrt{-g}
  J^i=\Omega_0\,\check{J}^i\,.
\end{equation}
Hence the gravitational field is involved in the inhomogeneous Maxwell
equation only via the metric dependent expression within the
parenthesis and via the determinant of the metric tensor.
Incidentally, since for the Reissner-Nordstr\"om solution the
determinant is the same as for a flat mani\-fold, the electromagnetic
field is not sensitive to the mass of the point charge.

In our axiomatic approach, see (\ref{maxwell1}), we arrive directly at
\begin{equation}\label{Max-eq''}
 \begin{cases} \hspace{3pt}
   dF=0\,,\\ \hspace{3pt} d\,^\star\! F=\Omega_0\,J\,,
 \end{cases}
\end{equation}
without discussing covariant derivatives. The Hodge star is known to
depend only on the conformal part of the metric, see Frankel
\cite{Ted}. If we identify the currents according to $J^i=\sqrt{-g}
\epsilon^{ijkl}J_{jkl}/6$, here $J=J_{ijk}\,dx^i\wedge dx^j\wedge
dx^k/6$ is the current 3-form, then the endresults of both procedures
coincide; compare (\ref{Max-eq'}),(\ref{Max-eq3}) with (\ref{Max-eq''}).

The Maxwell equations in tensor notation, if altered in the presence of
the gravitational field, can be rearranged by a change of the basic
notation.  In accordance with (\ref{spacetimerel}), the
electromagnetic excitation is linearly related to the field strength:
\begin{equation}\label{H-def}
  \check{H}^{ij}=\frac 12\,\chi^{ijkl}F_{kl}\,.
\end{equation}
The constitutive tensor reads
\begin{equation}\label{chiEM}
  \chi^{ijkl}(g)=\lambda_0\,\sqrt{-g}(g^{ik}g^{jl}-g^{il}g^{jk})\,.
\end{equation} 
Then the electromagnetic action (\ref{EM-act}) takes the form
\begin{eqnarray}\label{EM-act1}
  S(g,F)&=&-\frac 14\int F_{ij}\check{H}^{ij}d^4x= -\frac 12 \int
  F\wedge H\,,
\end{eqnarray}
whereas the field equations in tensor analytical and in exterior form
notations become, respectively,
\begin{equation}\label{EM-eq}
\begin{cases} \hspace{3pt}
  F_{ij,k}+F_{jk,i}+F_{ki,j}=0\,,\\ \hspace{3pt}
  \check{H}^{ij}{}_{,j}=\check{J}^i\,,
 \end{cases}
 \qquad {\rm or}\qquad
 \begin{cases} \hspace{4pt}
 dF=0\,,\\
 \hspace{3pt} dH=J\,.
 \end{cases}
\end{equation}
Consequently, in the Einstein-Maxwell system, the coupling is
completely determined by the constitutive tensor (\ref{chiEM}). This
coupling is referred to as {\em minimal coupling}.  Technically, the
minimal coupling is provided in the conventional coordinate dependent
textbook approach by the substitution of the partial by the covariant
derivatives (taken with respect to the Levi-Civita connection) and in
the axiomatic exterior calculus approach by Hodge staring the field
strength $F$.

Summing up we can state that the {\em minimal coupling\/} between
electrodynamics and gravity is provided by (\ref{maxwell1}) [or
(\ref{EM-eq}) with (\ref{H-def})]. The knowledge about the Riemannian
structure of spacetime is fed into the Hodge star operator via the
metric. In SR a flat metric appears, in GR a curved one.
\medskip

\noindent{\it Nonminimal coupling}\medskip

If for some reason we believe that the minimal coupling between
electrodynamics and gravity is insufficient and has to be generalized,
then we have to modify the action (\ref{EM-act}) by adding so-called
{nonminimal} terms. Still, we have to respect diffeomorphism and gauge
invariance.  Riemannian geometry is characterized by the curvature
tensor $R_{ijk}{}^l$ and its traces $ \hbox{Ric}_{ij}$ and $R$. Terms
proportional to the electromagnetic potential are forbidden because of
gauge invariance.  The term linear in $F$ vanishes as a product of a
symmetric and an antisymmetric tensors. Consequently the lowest order
nonminimal Lagrangian reads
\begin{equation}\label{FR-nonmin}
  L(F,R) = \a_1F_{ij}F^{kl}R^{ij}{}_{kl} +\a_2F_{ik}F^{jk}
  \hbox{Ric}^i{}_j +\a_3F_{ij}F^{ij}R\,.
\end{equation}
Let us recall several arguments in favor of a such non-minimal
couplings:\medskip

\noindent {(i)} For the values
$\a_1=\a_3=-4\a_2$, eq.(\ref{FR-nonmin}) can be recovered from the
5-dimensional Gauss-Bonnet action
\cite{Horndeski:1976gi,Horndeski:1977kz,Buchdahl:1979wi,
Mueller-Hoissen:1988bp}.\medskip

\noindent {(ii)} Another special set of parameters in (\ref{FR-nonmin})
\begin{equation}
  \a_1=-\frac{e^2\lambda_{\rm c}^2}{720\pi{\rm h}}\,,\quad
  \a_2=\frac{13e^2\lambda_{\rm c}^2}{720\pi{\rm h}}\,,\quad
  \a_3=-\frac{e^2\lambda_{\rm c}^2}{288\pi{\rm h}}\,,
\end{equation}
were derived in the one-loop approximation of quantum electrodynamics
(QED) \cite{Drummond:1979pp, Daniels:1993yi}.  Here $e$ is the
elementary charge, ${\rm h}=2\pi\hbar$ Planck's constant, and
$\lambda_{\rm c}=\hbar/(mc)$ the Compton wavelength of the
electron.\medskip

\noindent {(iii)} The Lagrangian  (\ref{FR-nonmin}) was used
as a basis for models with a variable speed of light
\cite{Novello:1989zb,Lafrance:1994in,Teyssandier:2003qh}.  Such an
effect is present in all models, except for the cases
$\a_1=\a_2=0$.\medskip

The constitutive tensor corresponding to the Lagrangian
$L(F,g)+ L(F,R)$ depends on the metric and the curvature tensor and its
contractions:
  \begin{eqnarray}\label{chiFR}
    \chi^{ijkl}(g,R)&=&(1-4\a_3
    R)(g^{ik}g^{jl}-g^{il}g^{jk})-4\a_1R^{ijkl} \nonumber\\ 
    &&-\a_2(g^{ik}\hbox{Ric}^{jl}-g^{il}\hbox{Ric}^{jk}+g^{jl}
    \hbox{Ric}^{ik}-g^{jk}\hbox{Ric}^{il})\,.
\end{eqnarray}
Since (\ref{chiFR}) is derived from a Lagrangian, its skewon part
(\ref{chi2}) is zero. Also the completely antisymmetric axion part
(\ref{chi3}) vanishes due to the symmetries of $R_{ijkl}$. Thus, in
(\ref{chiFR}) only the principal part ${}^{(1)}\chi^{ijkl} (g,R)$ is
left over.  Substituting (\ref{chiFR}) into the $3\times 3$
constitutive matrices (\ref{AB-matrix0}),(\ref{CD-matrix0}), we obtain
${\cal A}={\cal A}^T\,,{\cal B}={\cal B}^T\,,{\cal C}={\cal D}^T$.
The $\a^1$ and $\a^2$ terms yield birefringence in general.

%-----------------------------------
\subsection{Maxwell's field coupled to Einstein-Cartan gravity}
%-----------------------------------
In Einstein-Cartan gravity, see Blagojevi\'c \cite{Milutin}, e.g., the
spacetime geometry is of the Riemann-Cartan type and as such on two
fundamental structures --- the metric tensor $g_{ij}$ and the
connection $\Gamma_{ij}{}^k$. These quantities satisfy the metricity
condition
\begin{equation}
  g_{ij;k}=g_{ij,k} -\G_{ki}{}^l g_{lj} -\G_{kj}{}^l g_{il}=0\,.
\end{equation}

If the connection is not symmetric, then the torsion tensor
$T_{ij}{}^k$ is involved. In holonomic coordinates, it is defined by
$T_{ij}{}^k=2\Gamma_{[ij]}{}^k$. In exterior calculus, we have the
2-form $T^\a=D\vt^\a=T_{ij}{}^\a\,dx^i\wedge dx^j/2$, with the
exterior covariant derivative $D$.  Accordingly, a possible
interaction of torsion with the electromagnetic field is of special
interest \cite{RMP,Benn,Hammond,Puntigam,Shapiro,Hammond:rm,Preuss,
  gyros,torsion1}.

The {\it minimal\/} way of coupling the electromagnetic field with
Einstein-Cartan gravity works the same way as with Einstein gravity.
Since neither the exterior derivative $d$ nor the Hodge star $^\star$
can feel torsion directly, we have again the minimal equations
(\ref{maxwell1}):
\begin{equation}\label{aaaa}
  d\,^\star\!F = \Omega_0\,J\,,\qquad dF=0\,.
\end{equation}
They will do the job. 

The conventional coordinate dependent textbook approach led to
numerous misunderstandings. If one applies the semicolon rule to the
special relativistic formula $F_{ij}=A_{j,i}-A_{i,j}$, then one
arrives at
\begin{equation}\label{potEC}
  A_{j;i}-A_{i;j}=A_{j,i}-A_{i,j} -T_{ij}{}^kA_k=F_{ij}
  -T_{ij}{}^kA_k\,.
\end{equation}
This expression is diffeomorphism invariant. However, it is {\it
  not\/} gauge invariant under $A_i\rightarrow A_i+\partial_i\phi$.
Consequently, it has to be {\it rejected.} As we saw in the last
paragraph, the axiomatic exterior calculus approach leaves the
definition $F=dA$ intact, and there is no need, nor is it allowed, to
entertain in covariant derivatives in this context.

Accordingly, the {\em free\/} Maxwell Lagrangian is added to the
gravitational Lagrangian with the unchanged field strength.
Nevertheless, in an exact solution of a gravity-Maxwell system, the
torsion may depend on the electrical charge, as is exemplified by the
Reissner-Nordstr\"om solution with torsion in the framework of a
Poincar\'e gauge theory \cite{gron96}. In other words, torsion is
influenced by the electric charge {\em in\/}directly via the field
equations. What in the Lagrangian looks like ``no interaction at
all,'' still yields an effective ``minimal'' interaction.  \medskip

\noindent{\it Nonminimal coupling}\medskip

We are now looking for a complete family of nonminimally coupled
Maxwell-torsion Lagrangians that are diffeomorphism and gauge
invariant. Because of gauge invariance, the potential $A_i$ must not
appear in the action. Moreover, all expressions have to contain an
even number of indices in order, if contracted, to yield a scalar.
The expressions linear in the field strength are of the form
$\big(F\cdot (g^2\cdot T^2)\big)$.  Although such terms yield a family
of diffeomorphism and gauge invariant Lagrangians, they have to be
rejected from a physical point of view. Indeed, on the level of the
field equation, the Lagrangians linear in $F$ admit the existence of a
global electromagnetic field even without charges.  Such a
modification of classical electrodynamics seems to be unwarranted.

Hence the lowest order addenda to the Lagrangian are quadratic in
$F_{ij}$ and quadratic in torsion $T_{ij}{}^k$, i.e., of the typical
form $F^2\cdot g^3\cdot T^2$. The contractions can be performed in the
following three ways:\medskip

\noindent (i) All indices of the $F$-pair and of the $T$-pair are
contracted separately, $$\boxed{(F^2\cdot g^2) (g\cdot T^2)\,,}$$ or,
explicitly:
\begin{eqnarray}
\label{lag1} {}^{(1)}L(F,T)&=&F_{ij}F^{ij}\,
T_{mnk}T^{mnk}\,,\qquad \label{lag2}
{}^{(2)}L(F,T)=F_{ij}F^{ij}\,T_{mnk}T^{nkm}\,,\nonumber\\
\label{lag3}
{}^{(3)}L(F,T)&=&F_{ij}F^{ij}\,{T_{mn}}^n{T^{mk}}_k\,.
\end{eqnarray}

\noindent (ii) Two free indices of the $F$-pair are contracted
with two free indices of the $T$-pair, i.e., $$\boxed{(F^2\cdot
g)\cdot(g^2\cdot T^2)\,.}$$
Such terms are
\begin{eqnarray}
\label{lag4} {}^{(4)}L(F,T)&=&F_{im}{F_j}^m\,T^{inj}{T_{nk}}^k
\,,\qquad \label{lag5}
{}^{(5)}L(F,T)=F_{im}{F_j}^m\,T^{ikn}{T^j}_{kn}\,,\nonumber\\
\label{lag6} {}^{(6)}L(F,T)&=&F_{im}{F_j}^m \,T^{ikn}{T^j}_{nk}
\,,\qquad \label{lag7}
{}^{(7)}L(F,T)=F_{im}{F_j}^m \,T^{ikn}{T_{kn}}^j \,,\nonumber\\
\label{lag8}
 {}^{(8)}L(F,T)&=&F_{im}{F_j}^m\,{T^{ik}}_k{T^{jn}}_n \,.
\end{eqnarray}

\noindent (iii) The $F$-pair and the $T$-pair have four free
indices: $$\boxed{(F^2)\cdot (g^3\cdot T^2)\,,}$$
\begin{eqnarray}
 \label{lag9}
 {}^{(9)}L(F,T)&=&F_{ij}F_{kl}\,{T^{im}}_mT^{jkl} \,,\qquad
 \label{lag10}
 {}^{(10)}L(F,T)=F_{ij}F_{kl}\,{T^{im}}_mT^{klj} \,,\nonumber\\
 \label{lag11}
 {}^{(11)}L(F,T)&=&F_{ij}F_{kl}\,{T_m}^{ij}T^{mkl} \,,\qquad
 \label{lag12}
 {}^{(12)}L(F,T)=F_{ij}F_{kl}\,{T_m}^{ik}T^{mlj} \,,\nonumber\\
 \label{lag13}
 {}^{(13)}L(F,T)&=&F_{ij}F_{kl}\,{T_m}^{il}T^{mjk} \,,\qquad
 \label{lag14}
 {}^{(14)}L(F,T)=F_{ij}F_{kl}\,T^{mij}{T^{kl}}_m \nonumber\,,\\
 \label{lag15}
 {}^{(15)}L(F,T)&=&F_{ij}F_{kl}\,T^{mik}{T^{lj}}_m \,,\qquad
 \label{lag16}
 {}^{(16)}L(F,T)=F_{ij}F_{kl}\,T^{ijm}{T^{kl}}_m \,,\nonumber\\
 \label{lag17}
 {}^{(17)}L(F,T)&=&F_{ij}F_{kl}\,T^{ikm}{T^{lj}}_m \,.
\end{eqnarray}
Summing up, the general torsion Lagrangian  reads
\begin{equation}\label{gentor}
  \frac {1}{\ell^2}\,L(F,T)=-\frac 18\mathop{\sum}_{i=1}^{17}
  \b_{i}\,^{(i)}L(F,T)\,,
\end{equation}
with the dimensionless constants $\b_i$.

A special Lagrangian of type (\ref{gentor}), namely
\begin{equation}\label{eq4}
 ^\star(T_\a\wedge F)\, T^\a\wedge F=\frac 14\,
  \left(\,{}^{(1)}L-4\,{}^{(5)}L+\,{}^{(16)}L \right)\,^\star 1\,
\end{equation}
 was considered recently \cite {Solanki,Preuss} in the context
 of a cosmological
test of Einstein's equivalence principle.

Since we start with a modified Lagrangian, a skewon piece does not
 occur in the corresponding constitutive tensor. Thus
\begin{equation}\label{eq7}
\chi^{ijkl}(g,T)={\ell^2}\Big[\sum_{k\cdots
    q}\Big(T_{mnp}T_{qrs}\Big)\Big]^{[ij][kl]}=\widetilde{\chi}^{ijkl}(g,T)+
\check{\chi}^{ijkl}(g,T)\,,
 \end{equation}
 where $\widetilde{\chi}^{ijkl}(g,T)$ is the modification of the
 principal part, while the axion part takes the form
\begin{equation}\label{eq8}
  \check{\chi}^{ijkl}(g,T)=\ell^2\Big[\sum_{m\cdots
    s}\Big(T_{mnp}T_{qrs}\Big)\Big]^{[ijkl]}\,.
\end{equation}

\noindent{\bf Birefringence   induced by torsion. }
For the Lagrangians (\ref{lag1})--(\ref{lag17}) we find the following
effects on $\widetilde{\chi}^{ijkl}$ and on the light cone (we put
$\ell=1$):\medskip

\noindent (i) The Lagrangians (\ref{lag1}) yield
\begin{equation}\label{eq10a}
  \widetilde{\chi}^{ijkl}(g,T)=S(g^{ik}g^{jl}-g^{il}g^{jk})=
  2Sg^{i[k}g^{l]j}\,,
\end{equation}
where $S$ is a scalar function quadratic in torsion. Therefore,
the Tamm-Rubilar tensor density changes only by a conformal
factor and the
light cone is preserved. %\medskip

\noindent (ii) For (\ref{lag4}), we introduce the abbreviation
$S^{ij}=S^{ji}:=\big[g^2\cdot T^2\big]^{(ij)}$. The axion field
is absent. Thus,
\begin{equation}\label{eq10b}
   \widetilde{\chi}^{ijkl}(g,T)%&=&\frac
 % 14\left(g^{ac}S^{bd}-g^{ad}S^{bc}+g^{bd}S^{ac}-g^{bc}S^{ad}\right)
 % \nonumber\\ &=&
=g^{[i|[k}\,S^{l]|j]}\,.
\end{equation}
The $3\times 3$ constitutive matrices
${\cal A,B,C,D}$ read:
\begin{eqnarray}\label{eq10c}
  A^{ab}(g,T)&:= &\widetilde{\chi}^{0b0a}(g,T)=\frac 14
  (g^{00}S^{ab}+g^{ab}S^{00})\,,\nonumber\\
   B_{ab}(g,T)&:=&\frac
  14\epsilon_{bcd}\epsilon_{aef} \,
  \widetilde{\chi}^{cdef}(g,T)=\frac
  14\left(g_{ab}S_c{}^c-S_{ab}\right)\,,\nonumber\\
 {C^a}_b(g,T)&:=&\frac 12\epsilon_{bcd}\,
 \widetilde{\chi}^{cd 0a}(g,T)=\frac 14 {\epsilon^a{}_{bc}}\,S^{0c}.
\end{eqnarray}
These matrices, with $ {C^a}_b(g,T)={D_b}^a(g,T)$, obey
\begin{equation}\label{eq10d}
\hspace{-3pt} A(g,T)= A^{\rm T}(g,T),\,\quad
  B(g,T)= B^{\rm T}(g,T),\,\quad C(g,T)=D^{\rm T}(g,T)
\end{equation}
(${\rm T}$ = transposed). These relations, together with the
closure condition for $\widetilde{\chi}^{ijkl}(g,T)$, guarantee
the uniqueness of the light cone \cite{Birkbook}. Thus
birefringence does not emerges in this group of models
either.\medskip

\noindent (iii) For the Lagrangians from the third group
(\ref{lag9}), birefringence is a generic property. This was shown
for (\ref{eq4})  in the case of spherically symmetric torsion
 \cite{Preuss,torsion1}.
\medskip

\noindent{\bf  Axion field induced by torsion.} 
The axion, that is, a pseudo-scalar field, was extensively studied in
different contexts of field theory. Its quantized version is believed
to provide a solution to the strong CP problem of quantum
chromodynamics (QCD). The emergence of axions is a general phenomenon
in superstring theory.  The axion as a classical field also appears in
various discussions of the equivalence principle in gravitational
physics and in inflationary models. In the context discussed in this
paper, the axion field is introduced in (\ref{chi3}) with
(\ref{Salpha})$_2$ as one irreducible piece of the electromagnetic
constitutive tensor of spacetime.  According to (\ref{eq8}),
we may imagine that in the present case the axion field is induced by
the nonminimal coupling of the Maxwell field to the torsion of
spacetime. Explicitly, we find the following:\medskip

\noindent (i) The Lagrangians (\ref{lag1}) yield
\begin{equation}\label{eq8xxx}
 \check{\chi}^{ijkl}(g,T)= \ell^2g^{[ik}g^{jl]}(T\cdot T)=0\,.
\end{equation}%\medskip

\noindent (ii) The Lagrangians (\ref{lag4}) yield
\begin{equation}\label{eq8xx}
  \check{\chi}^{ijkl}(g,T)= \ell^2 g^{[jl}(T\cdot T)^{ik]}=0\,.
\end{equation}%\medskip

\noindent (iii) For those of the third group (\ref{lag9}), the
general form of the axion field
$\alpha\!=\!\epsilon_{ijkl}\,\check{\chi} ^{ijkl}(g,T) /4!$ is
\begin{equation}\label{eq9}
  \alpha=\ell^2\epsilon_{ijkl}\Big(
  \a_1{T^{ijm}}{T^{kl}{}_m}+\a_2T^{ijk}{T^{lm}}_m\Big)\,,
\end{equation}
where $\a_1,\a_2$ are free dimensionless parameters, which are linear
combinations of the $\b$'s. For the special case (\ref{eq4}), the
axion is extracted from (\ref{eq9}) by putting $\a_1=1,\a_2=0$.

If in vacuum an axion field $\alpha$ emerges, then the Lagrangian
picks up an additional piece $\sim\alpha\,F\wedge F$, see
\cite{Birkbook}.  Accordingly, the Maxwell equations are those
displayed in (\ref{MaxAx}). Only a non-constant axion field contributes.
The coupling of the Maxwell field to a non-constant axion field, in
the case of a plane electromagnetic wave, amounts to a rotation of the
polarization vector of the wave, see \cite{clausannalen}, i.e., the
axion field induces an {\it optical activity\/}.  \medskip

%-----------------------------------
\subsection{Maxwell's field coupled to metric-affine  gravity}
%-----------------------------------
Metric-affine gravity (MAG) is based on a spacetime geometry with
completely independent metric and connection \cite{Hehl:1994ue}. Thus,
in addition to the torsion tensor, the nonmetricity tensor emerges
\begin{equation}\label{nonmetr}
  Q_{kij}=-g_{ij;k}=-g_{ij,k} +\G_{ki}{}^l g_{lj} +\G_{kj}{}^l
  g_{il}\,.
\end{equation}
In exterior calculus, we have the nonmetricity 1-form
$Q_{\a\b}:=-Dg_{\a\b}$, with the decomposition
$Q_{\a\b}=Q_{i\a\b}dx^i$. In this framework, we can look for a
possible interaction of the nonmetricity with the electromagnetic
field. As in Einstein and Einstein-Cartan gravity, the minimal
coupling remains untouched, that is, we have again
$\lambda_0\,d\,^\star F=J,\;dF=0$. 

In {\it nonminimal coupling,} however, there emerge two additional
types of terms, namely $F^2\cdot g^{5}\cdot Q^2$ and $F^2\cdot
g^4\cdot Q\cdot T$.  These coupling terms can be expanded in the
following ways:\medskip

\noindent (i) All indices of the $F$-pair and of the $Q$-pair (or
of the $TQ$-pair) are contracted separately: $$\boxed{(F^2\cdot
g^2) (g^3\cdot Q^2)\,,} \qquad {\rm or}\qquad \boxed{(F^2\cdot g^2)
(g^2\cdot Q\cdot T)\,.}$$
%, or explicitly:
%{}^{(1)}L(F,Q)&=&
% F_{ab}F^{ab}\,Q_{mnk}Q^{mnk}\,,\qquad %\label{lagQ2}
%{}^{(2)}L(F,Q)=F_{ab}F^{ab}\,Q_{mnk}Q^{mkn}\,,\nonumber\\
%\label{lagQ3}
% {}^{(3)}L(F,Q)&=&F_{ab}F^{ab}\,{Q_{mn}}^n{Q^{mk}}_k\,,\qquad
% {}^{(4)}L(F,Q)=F_{ab}F^{ab}\,Q^m{}_{mn}Q^{nk}{}_k\,,\nonumber\\
% {}^{(5)}L(F,Q)&=&F_{ab}F^{ab}\,Q^m{}_{mn}Q^{nk}{}_k\,.
%\end{eqnarray}
Examples of such terms can be easily constructed:
\begin{equation}\label{lagQ1}
{}^{(1)}L(F,Q)=F_{ij}F^{ij}\,Q_{mnk}Q^{mnk}\,, \quad
{}^{(1)}L(F,Q,T)=F_{ij}F^{ij}\,Q_{mnk}T^{mnk}\,.
%{}^{(1)}\chi^{abcd}(g,Q)=(g^{ac}g^{bd}-g^{ad}g^{bc})\,Q_{mnk}Q^{mnk}\,.
\end{equation}
The corresponding   constitutive tensors are of the form ($S$ is
a scalar)
\begin{equation}\label{chiQ1}
\chi^{ijkl}(g,Q)=(g^{ik}g^{jl}-g^{il}g^{jk})S\,.
\end{equation}
Certainly the axion field and the birefringence effect are absent in this
group of models.\medskip

\noindent (ii) Two free indices of the $F$-pair are contracted
with two free indices of the $Q$-pair (or $TQ$-pair), i.e.,
$$\boxed{(F^2\cdot g)\cdot(g^4\cdot Q^2)\,,} \qquad {\rm {or}}\qquad
\boxed{(F^2\cdot g)\cdot(g^3\cdot Q\cdot T)\,.}$$ Examples of such
Lagrangians are
\begin{equation}\label{lagQ2}
{}^{(2)}L(F,Q)=F_{ik}F_j{}^{k}\,Q^m{}_{mn}Q^{ijn}\,, \quad
{}^{(2)}L(F,T,Q)=F_{ik}F_j{}^{k}\,T_{mn}{}^nQ^{imj}\,.
\end{equation}
The corresponding constitutive tensor is of the form
\begin{equation}\label{chiQ2}
\chi^{ijkl}(g,Q)=g^{[i|[k}\,S^{l]|j]}\,,
\end{equation}
where $S^{ij}=(g^4\cdot Q^2)^{ij}$ or $S^{ij}=(g^3\cdot Q\cdot T
)^{ij}$. This tensor is necessary symmetric since it multiplies
the symmetric tensor $F_{ik}F_j{}^{k}$. Similarly to the
non-minimal coupling to torsion (\ref{eq10b}-\ref{eq10d}), the
birefringence effect is absent in this group of models. Comparing
to (\ref{eq8xx}) we see that also the axion part of this
constitutive tensor is zero.\medskip

\noindent (iii) The $F$-pair and the $Q$-pair (or $TQ$-pair) have
four free indices:
$$\boxed{(F^2)\cdot (g^5\cdot Q^2)\,,}\qquad{\rm {or}}\qquad
\boxed{(F^2)\cdot (g^4\cdot T\cdot Q)\,.}$$ For instance,
\begin{equation}\label{lagQ3}
{}^{(3)}L(F,Q)=F_{ij}F_{kl}\,Q^{ik}{}_mQ^{jlm}\,, \quad
{}^{(3)}L(F,T,Q)=F_{ij}F_{kl}\,T^{im}{}_mQ^{kl}{}_i\,.
\end{equation}
Possessing an axion field and birefringence are generic properties in
this group of models.

In MAG, all the non-minimal coupling terms described above are of the
same quadratic order in the connection. Therefore, additional
Lagrangians depending on curvature, torsion, and nonmetricity have in
general to be considered together.

%%%%%%%%%%%%%%%%%%%%%%%%%%%%%%%%%%%%%%%%%%%%%%%%%%%%%%%%%
\section{Discussion}
%%%%%%%%%%%%%%%%%%%%%%%%%%%%%%%%%%%%%%%%%%%%%%%%%%%%%%%%%

Although Maxwell's electrodynamics is a firmly established classical
field theory, it is still open to new developments along many lines.
Furthermore, the study of the fundamental structures of Maxwell's
theory can serve as a natural starting point for non-abelian
modifications with applications in high-energy physics and gravity. In
particular, modern string and brane theories rely heavily on
classical field-theoretic structures.  In addition, there are
important classical problems still awaiting for their solutions, such
as the quantization of the electric charge, the existence of magnetic
monopoles, and the value of the coupling constant, as well as a number
of new open issues: (i) The physical place and the role of the
premetric partners of the standard electromagnetic field, such as the
axion, the skewon, and the dilaton fields. (ii) The appropriate form
of the coupling between the electromagnetic and the gravitational
field (also taking into account the possible gauge-theoretic
extensions of standard GR, such as Poincar\'e gravity and, more
generally, metric-affine gravity). (iii) The description of
a high-energy electromagnetic field in a material medium.

In this lecture we gave an overview of the recent developments of the
axiomatic premetric approach based on the conservation of electric
charge and of magnetic flux as well as on some additional inputs
such as the structure of the energy-momentum. These facts are well
established theoretically and tested experimentally. The field
variables and the corresponding field equations are then
straightforwardly derivable from the basic axioms when certain natural
restrictions on the topology of spacetime are assumed. The resulting
construction is actually a topological one, in the sense that it does
not depend on a specific geometry of the underlying spacetime
manifold. Only the additional notion of a constitutive relation brings
in the information on the specific structure of the geometry of
spacetime. The open problems, indicated above, are directly linked to
the investigation of the properties of the constitutive relation.
Going beyond the linear and local constitutive law is, in fact, a step
which takes into account the quantum nature of the actual matter
sources. The corresponding dynamics of the electromagnetic field
becomes highly nontrivial. A good recent analysis of the earlier
nonlinear electrodynamical models can be found in
\cite{DelphenichNonlinear}; specific results on the birefringence in
such theories were obtained in \cite{Yu+GuNonlinear}.

\medspace\medspace
 \noindent{\bf Acknowledgments:} One of the authors would
 like to thank Branko Drako\-vich, Milutin Blagojevi\'c, and Djordje
 \v Sija\v cki for the invitation to give lectures in Zlatibor and for
 their hospitality. He also thanks Milutin Blagojevi\'c for many
 interesting discussions.

\end{document}